\documentclass[a4paper,aps,prd,twocolumn,showkeys,showpacs,nofootinbib]{revtex4}
\usepackage{latexsym}
\usepackage{amsbsy}
\usepackage{mathrsfs}
\usepackage{color}
\usepackage{psfrag}
\usepackage{enumerate}
\usepackage{amsmath,amssymb,calc,amsfonts}
\usepackage{graphicx,calc,epsfig}
\usepackage[multiple]{footmisc}
\usepackage[english]{babel} 
\usepackage{mathtools}
\usepackage{bbm}
\usepackage{accents}
\usepackage{dsfont}

\def\ut#1{\rlap{\lower1ex\hbox{$\sim$}}#1{}}

\newcommand{\be}{\nopagebreak[3]\begin{equation}}
\newcommand{\ee}{\end{equation}}
\newcommand{\ba}{\nopagebreak[3]\begin{eqnarray}}
\newcommand{\ea}{\end{eqnarray}}
\DeclareFontFamily{U}{rsfs}{}         
\DeclareFontShape{U}{rsfs}{m}{n}{<5> rsfs5 <6><7> rsfs7          
  <8><9><10><10.95><12><14.4><17.28><20.74><24.88> rsfs10}{}     
\DeclareMathAlphabet{\mathfs}{U}{rsfs}{m}{n}                     
                               
\def\i{i}

\def\pb#1{\rlap{\lower1.5ex\hbox{$\longleftarrow$}}{#1}}
\def\dpb#1{\rlap{\lower1.5ex\hbox{$\Longleftarrow$}}{#1}}
\def\spb#1{\rlap{\lower1.5ex\hbox{$\leftarrow$}}{#1}}
\def\sdpb#1{\rlap{\lower1.5ex\hbox{$\Leftarrow$}}{#1}}

\definecolor{blue}{rgb}{0,0,1}
\definecolor{green}{rgb}{0,1,0}
\definecolor{red}{rgb}{1,0,0}
\definecolor{vio}{rgb}{1,0,1}
\definecolor{ama}{rgb}{1,1,0}
\usepackage[all]{xy}
\xyoption{knot}
\xyoption{frame}
\xyoption{arc}
\xyoption{curve}
\xyoption{poly}
\usepackage{hyperref}
\hypersetup{
  colorlinks   = true, 
  urlcolor     = blue, 
  linkcolor    = blue, 
  citecolor   =  red 
}
\newtheorem{Definition}{Definition}[section]

\def\be{\begin{equation}}
\def\ee{\end{equation}}
\def\ba{\begin{eqnarray}}
\def\ea{\end{eqnarray}}

\begin{document}

\title{Anyonic statistics and large horizon diffeomorphisms for\linebreak Loop Quantum Gravity black holes}

\date{\today}

\author{Andreas G. A. Pithis}\email{andreas.pithis@kcl.ac.uk}
\affiliation{Department of Physics, King's College London, University of London, Strand, London, WC2R 2LS, U.K., EU}
\author{Hans-Christian Ruiz Euler}\email{Hruiz@science.ru.nl}
\affiliation{Donders Institute for Brain Cognition and Behaviour, Radboud University Nijmegen, 6525 EZ Nijmegen, The Netherlands, EU}

\begin{abstract}
We investigate the role played by \textit{large} diffeomorphisms of quantum isolated horizons for the statistics of Loop Quantum Gravity (LQG) black holes by means of their relation to the braid group. To this aim the symmetries of Chern-Simons theory are recapitulated with particular regard to the aforementioned type of diffeomorphisms. For the punctured spherical horizon, these are elements of the mapping class group of $S^2$, which is almost isomorphic to a corresponding braid group on this particular manifold. The mutual exchange of quantum entities in two dimensions is achieved by the braid group, rendering the statistics anyonic. With this we argue that the quantum isolated horizon model of LQG based on $SU(2)_k$-Chern-Simons theory exhibits non-abelian anyonic statistics. In this way a connection to the theory behind the fractional quantum Hall effect and that of topological quantum computation is established, where non-abelian anyons play a significant role.
\end{abstract}



\maketitle

\section{Introduction}

When considering spacetimes with boundaries or asymptotic regions, the variational principle applied to the Einstein-Hilbert action is solely fulfilled upon the imposition of boundary conditions onto the fields and the inclusion of boundary terms into the action \cite{YGH}. As a consequence of the latter, gauge and diffeomorphism invariances get broken and field configurations which used to be in the same gauge orbit are not so anymore. Differently spelled, at the boundary gauge transformations become symmetry transformations \cite{Carlip} and the physical states of the respective quantum theory are allowed to transform under representations of the group of these boundary transformations \cite{RT}.\newline 
Distinct from them are \textit{large} gauge and diffeomorphism transformations which are not imposed by constraints but also have an interesting effect on the boundary states of the theory \cite{TeitelboimHenneaux}. This motivates us to have a closer look onto the taxonomy of transformations in the Loop Quantum Gravity (LQG) description of black holes based on isolated horizons (IH) and Chern-Simons (CS) theory.\newline
The horizon of black holes as an inner boundary of space can be described in equilibrium locally by the isolated horizon boundary condition \cite{IsolatedHorizonFramework}. The introduction of this notion is justified since the usual definition of a black hole as a spacetime region of no escape is global. This means that it requires the knowledge of the entire spacetime as well as that it be in equilibrium and consequently it does not appear to be useful for the description of local physics. However, these problems are solved within the quasilocal notion of an isolated horizon. From a physical point of view its introduction amounts to having no fluxes of matter and/or gravitational energy across it. From a technical point of view the boundary conditions lead to a surface term for the horizon in the overall action of the gravitational field which in terms of Ashtekar-Barbero variables is proportional to the action of a topological gauge theory, namely CS-theory. Furthermore, one can show that this description is fully compatible with the laws of black hole mechanics.\newline 
The quantum geometric handling of spacetimes with such an isolated horizon by means of LQG techniques describes the quantum geometry of the bulk by a spin network, whose graph pierces the horizon surface yielding punctures. The totality of the punctures forms a gas of topological defects which represent the quantum excitations of the gravitational field of the horizon. These black hole quantum d.o.f. are then described by $SU(2)$-CS-theory at level $k$ given on a punctured $2d$-sphere \cite{ABK,SU2}.\newline
Equipped with this, one sets out to count the microstates of the corresponding Hilbert space \cite{KaulMajumdar,SU2Character,ABK,BE,PP,PranzettiPolo,Me,GM,SU2,SU2rev}. Together with the introduction of proper notions of a quasi-local energy and a local temperature of the isolated horizon its statistical mechanical analysis is facilitated \cite{APE, espera, Pranzetti}. With this an expression for the entropy is obtained which is remarkably compatible with the semiclassical Bekenstein-Hawking area law \cite{Bekenstein, Hawking} up to a quantum hair correction due to the quantum geometry of the isolated horizon.\newline
Despite these successes in matching the semiclassical results, the question was raised whether the statistics of the quantum gravitational d.o.f. could actually be different \cite{espera, Sahlmann} from the one assumed, considering the well-known fact from solid state physics that quantum objects in $2d$ obey anyonic statistics. This is the reason why we are motivated to turn hereafter to the seemingly more exotic type of statistics, generally termed as anyonic/braiding statistics and investigate its bearing on the LQG black hole model. Drawing from CFT techniques, we will be led to the reinterpretation of the quantum isolated horizon model as one which explicitly exhibits non-abelian anyonic statistics.\newline
To this aim, the article is organized as follows. As a background for our work we assume the isolated horizon framework \cite{IsolatedHorizonFramework} and its quantization \`{a} la LQG. Since there one borrows techniques from CS-theory \cite{ABK,SU2}, for reasons of self-consistency and completeness we will firstly review the symmetries of CS-theory as well as its Hamiltonian formulation in section (\ref{subsectionA}). Then we recapitulate properties of the LQG black hole model and its statistics in the following subsection (\ref{subsectionB}). Assuming that in the purely gravitational case the horizon states are distinguishable throughout the article, we will summarize the motivation for this in (\ref{subsectionC}). The reader familiar with these reviewed concepts and ideas is invited to jump directly into the third and core section (\ref{SectionIII}) where we elaborate the main and new results. There we will firstly inspect the topological features of the physical phase space and its relation to the braid group, which reveals the anyonic nature of the horizon degrees of freedom in (\ref{subsectionA2}). Afterwards we will further investigate the braid group symmetry of the punctured sphere by relating it to the large diffeomorphisms of the horizon and discuss the effect of the occurring non-abelian phases in (\ref{subsectionB2}). We will then connect the discussion of this property of the horizon d.o.f. to formal aspects of the theory of non-abelian anyons known from solid state physics in (\ref{subsectionC2}). Since the article suggests that the braiding symmetry/ statistics is suppressed for a large values of the CS-level $k$, we comment on the sensitivity of the entropy to $k$ and give qualitative arguments why the black hole radiance spectrum should display traces of the braiding in (\ref{subsectionD2}). Finally, the last section (\ref{SectionIV}.) closes the article with a discussion of the results and comments on possible future investigations. The Appendixes supplement the material where suitable and needed.

\section{Chern-Simons theory and LQG black holes}\label{SectionII}
\subsection{Symmetries of Chern-Simons theory}\label{subsectionA}
Within this subsection we will revise essentials of CS-theory with special regard to its symmetry properties, the difference between small and large diffeomorphisms and its Hamiltonian formulation which will be exploited afterwards.\newline
The action of CS-theory on an oriented smooth $3$-manifold $M$ is given as
\be\label{CS-WirkungaufM}
S_{CS}[\tilde{A}]=\frac{k}{4\pi}\int_M {\rm Tr}(d\tilde{A}\wedge \tilde{A}+\frac{2}{3}\tilde{A}\wedge \tilde{A}\wedge \tilde{A}),
\ee
for a $G$-valued connection $\tilde{A}=\tilde{A}_{\mu}^iJ_i dx^{\mu}$ and $k$ denotes the coupling constant (level). $G$ is a compact, simple and simply connected Lie group and the generators $\{J_i\}$ with $i=1...\dim G$ form the basis of the corresponding Lie algebra. Stationarity of the action leads to the equation of motion 
\be\label{ELeqnOSCond}
F=d\tilde{A}+\tilde{A}\wedge \tilde{A}=0.
\ee
Inspecting its gauge symmetries, the \textit{overall} gauge group is given by the semi-direct product of ${\rm Diff}_0(M)$ with the infinite dimensional and possibly topologically non-trivial $\mathcal{G}=C^{\infty}(M,G)$ \cite{Witten, Zuckerman}.\newline
Let us dwell for a moment on this point and firstly consider transformations which are elements in $\mathcal{G}$. This leads us to the well-known transformation law for the connection 
\be\label{GTforConlaw}
\tilde{A}\to \tilde{A}^g=g\tilde{A}g^{-1}-(dg)g^{-1},
\ee 
with $g\in\mathcal{G}$. In fact, $\mathcal{G}$ comprises two parts which are called small and large gauge transformations. We call gauge transformations small if they are connected to the identity and one easily sees that $S_{CS}$ is invariant with respect to them. Let $g$ be such a transformation given in its finite form as $g=\exp(i J_i \zeta^i)$ where $\zeta^i$ are the gauge parameters. Infinitesimally, $g\approx 1-i J_i\zeta^i$ with $\zeta<<1$ and this yields 
\be\label{Ainfgauge} 
\delta \tilde{A}=\tilde{A}^g-\tilde{A}\approx d_{\tilde{A}}\zeta.
\ee
Importantly, invariance under small gauge transformations is not enough to guarantee the invariance with respect to finite transformations. This is due to the fact that there are topologically non-trivial finite gauge transformations with homotopy class different from $0$. One calls them large gauge transformations. If one demands that the path integral
\be\label{CSPI}
Z_k(M)=\int \mathcal{D}\tilde{A}~e^{iS_{CS}[\tilde{A}]}
\ee
is a gauge invariant object with respect to small and large gauge transformations, it can be shown that for closed $M$ and compact $G$ the coupling constant $k$ must be an integer and is hence discrete.\newline 
Similar to $\mathcal{G}$, one differentiates between two types of diffeomorphisms, namely small and large ones. Diffeomorphisms in ${\rm Diff}_0(M)$ are homotopic to the identity, can be infinitesimally generated and are called small. Since CS-theory is a TQFT of Schwarz type, its action, equations of motion and observables do not require the existence of a metric. It is thus diffeomorphism invariant \cite{Witten, Labastida}, i.e. invariant with respect to ${\rm Diff}_0(M)$. 
Large diffeomorphisms on the other hand cannot be obtained from summing up an infinite number of infinitesimal transformations and are not homotopic to the identity. They form a group called the mapping class group which is denoted by ${\rm MCG}(G)={\rm Diff}(M)/{\rm Diff}_0(M)$.\newline
In the context of CS-theory one can show that on shell, i.e. when (\ref{ELeqnOSCond}) is fulfilled, small diffeomorphisms are equivalent to small gauge transformations. To see this, consider the change of the connection $\tilde{A}$ under an infinitesimal coordinate transformation $x^{\mu}\to x^{\mu}+\xi^{\mu}$ with $\mu\in\{0,1,2\}$. This is expressed as
\be\label{Ainfdiffeo}
\delta_{\xi}\tilde{A}=L_{\xi}\tilde{A}=(i_{\xi}d+d i_{\xi})\tilde{A}=i_{\xi} F +d_{\tilde{A}}(i_{\xi}\tilde{A}),
\ee
wherein $d_{\tilde{A}}$ denotes the gauge-covariant exterior derivative and $\xi$ is an infinitesimal generator of small diffeomorphisms. On shell this expression is just an ordinary infinitesimal gauge transformation (\ref{Ainfgauge}) with the gauge parameter $\zeta^{i} = \xi^{\mu}\tilde{A}_{\mu}^i$.\newline
In stark contrast to this, large diffeomorphisms and large gauge transformations are discrete and strictly distinct symmetries of the theory. In the quantum theory one cannot simply demand that quantum states should be invariant under the action of these groups. Instead, they can act as symmetry transformations on the states. In the later course of this article we will argue for the importance of the mapping class group for the treatment of the quantum isolated horizon framework of LQG and we will relate it to the statistical symmetry giving rise to braided/anyonic statistics.\newline
All these symmetry considerations also hold for the Hamiltonian formulation of CS-theory \cite{Witten, Guadagnini, AlekseevSchomerus, AGN}. There, the gauge field is split into $\tilde{A} =A_0 dx^0+A_a dx^a$ due to the product structure of $M=\mathbb{R}\times\Sigma$, where $\Sigma$ is an arbitrary orientable surface. Then the spatial components $A=A_a dx^a$ of the gauge field are considered as the dynamical variables. The appearing $A_0$-component has null conjugate momentum and serves as a Lagrange multiplier in the action
\be\label{2plus1splitaction}
S=\frac{k}{4\pi}\int_{\mathbb{R}}\int_{\Sigma}{\rm Tr}(-A\partial_0 A+2A_0 F),
\ee
enforcing the first class constraint $F=0$. From the infinitesimal variation of the action one also obtains a boundary term, which we can identify as the symplectic potential
\be\label{CSsymp1form}
\theta=\frac{k}{4\pi}\int_{\Sigma}{\rm Tr}(A\wedge\delta A)+\delta\rho[A].
\ee
Therein $\rho$ denotes an arbitrary functional of $A$ and $\delta\rho$ expresses the freedom of canonical transformations \cite{Nair}. The symbol $\delta$ corresponds to the exterior derivative on the space of gauge potentials on $\Sigma$. With this the symplectic $2$-form is obtained by
\be\label{CSsymp2form}
\omega=\delta \theta=\frac{k}{4\pi}\int_{\Sigma}{\rm Tr}(\delta A\wedge\delta A).
\ee
If gauge symmetries have not yet been reduced out, $\omega$ is presymplectic and thus has zero modes generating gauge symmetries as discussed above. Upon symplectic reduction one yields the physical or reduced phase space. We consider $\omega$ to be non-degenerate below.\newline
Together with the physical phase space given by the moduli space of flat connections
\be\label{phasespacewithoutpunctures}
\Gamma=\{A| F=0\}/\mathcal{G},
\ee
we have a symplectic manifold $(\Gamma,\omega)$, where $\mathcal{G}=C^{\infty}(\Sigma,G)$.

\subsection{LQG black hole model and its statistical mechanics}\label{subsectionB}
This subsection revises parts of the classical isolated horizon framework and its quantization \cite{ABK, SU2, PP, PranzettiPolo} and summarizes essentials of their statistical mechanical analysis as in \cite{espera}. This review material is needed for the understanding of the rest of this article in view of the anyonic statistics to be analyzed in section (\ref{SectionIII}).\newline
The isolated horizon field theory lives on a $3$-manifold $\Delta$, which is a cylinder $\Delta =\mathbb{R}\times S^2$, where $\mathbb{R}$ parametrizes the time $t$ and $G$ is $SU(2)$ hereafter. Spherically symmetric isolated horizons can be described as a dynamical system by a presymplectic form $\omega_{\rm horizon}$, which corresponds to that of an $SU(2)$-CS-theory. For a proof and a general discussion see \cite{ABK, SU2, PP,PranzettiPolo}. Physically this means that the gravitational field of the horizon resides in a topological phase. The overall symplectic structure splits as
\be\label{overallsympstr}
\omega_{\rm total}=\omega_{\rm bulk}+\omega_{\rm horizon}
\ee
and field components from bulk and horizon are coupled properly together by the IH boundary condition which in terms of Ashtekar-Barbero variables reads as
\be\label{IHBC}
F^i(A^i)+\frac{\pi(1-\gamma^2)}{a_H}\Sigma^i=0,
\ee
where $a_H$ denotes the classical horizon area and $\gamma$ is the Immirzi parameter. ${F}^i$ is the curvature $2$-form of the Ashtekar-Barbero connection $A^i$ being pulled back to $S^2$ and $\Sigma^i$ denotes the solder $2$-form of the bulk theory and the internal index $i\in\{1,2,3\}$ indicates that the respective object is colored with an element of $su(2)$ in the defining representation.\newline
In the following we will use that in LQG one regularizes the Poisson algebra of the Ashtekar-Barbero connection and the densitized triad. To this aim, one smears the connection along a path yielding a holonomy and the densitized triad along a surface yielding the flux, respectively. The resulting smeared algebra is the so-called holonomy-flux algebra \cite{LQG}. If one embeds a surface of spherical topology such as the one of a classical isolated horizon $\Delta$ into a surrounding spacelike $3$-space, then it will be pierced by paths $\gamma$ (supporting the bulk holonomies) at the points $\mathcal{P}=\{p_1,...,p_N\}$. One interprets this set as a distribution of sources on $S^2$, each labeled with a representation $\{\rho_p\}_1^N$ of $su(2)$ and $\Sigma^i$ is then
\be
\Sigma^i=16\pi\ell_p^2\gamma\sum_{p\in\mathcal{P}} J_{\rho_p}^i\delta^2(x,x_p).
\ee
Substituting this into (\ref{IHBC}) leads to
\be
F^i+\frac{4\pi}{k}\sum_{p\in\mathcal{P}} J_{\rho_p}^i\delta^2(x,x_p)=0
\ee
and the action for the horizon theory using (\ref{2plus1splitaction}) is thus
\be\label{ActionPunctures}
S_{\rm horizon}=\text{eq.} (\ref{2plus1splitaction})+\int_\mathbb{R}dt~{\rm Tr}\bigl(\sum_{p\in\mathcal{P}} J_{\rho_p}^i A_{0i}(x_p)\bigr),
\ee
where the level of the CS-theory is now given by $k=a_H/(4\pi\gamma(1-\gamma^2)\ell_p^2)$ and $\ell_p$ is the Planck length. The Euler-Lagrange equations lead to the (induced) constraint 
\be\label{curvaturewithpunctures}
G^i\equiv F^i+\frac{4\pi}{k}\sum_{p\in\mathcal{P}} J_{\rho_p}^i\delta^2(x,x_p)=0,
\ee
delineating that the curvature of the connection on the surface is concentrated at the points of the punctures/defects. This first class constraint generates (small) gauge transformations and (small) diffeomorphisms. More precisely, the horizon part of the smeared Gauss constraint is
\be\label{GaussConstraint}
G[\zeta,A]=\int_{S^2} \zeta_i G^i\approx 0,
\ee
for all $\zeta:\Delta\to su(2)$, whereas the diffeomorphism constraint is
\be\label{VectorConstraint}
V[\xi,A]=\int_{S^2}\xi^{\mu}A_{\mu i} G^i\approx 0,
\ee
for all vectors $\xi$ $(\mu=\theta,\phi)$ which are tangent to the horizon. The form of (\ref{GaussConstraint},\ref{VectorConstraint}) implies the on-shell equivalence of small diffeomorphisms and small gauge transformations as in (\ref{Ainfgauge},\ref{Ainfdiffeo}). In addition, for the sources at the points $\{x_p\}_1^N$ one has conjugations
\be
J^{i}_{\rho_{p}}\to J^{i g}_{\rho_{p}}=g^{-1}J^{i}_{\rho_{p}}g\in\mathcal{C}_{p}^g
\ee
and the gauge invariance of $F^i$ implies $\mathcal{C}_p=\mathcal{C}_p^g$ \cite{AlekseevSchomerus,AGN}.\newline
The physical phase space of this system is then given as
\be\label{phasespacepuncturessm}
\Gamma=\{\{A|F=0 \}\times\mathcal{C}_{1}\times\ldots\times\mathcal{C}_{N}\}/\{\text{gauge transformations}\}
\ee
as in \cite{AlekseevSchomerus,AGN}.\newline
The form of the overall symplectic structure (\ref{overallsympstr}) motivates us to quantize the bulk and horizon d.o.f. separately. The quantum geometry of the bulk is given by a spin network, whose graph impinges on the horizon surface yielding the punctures. Hence, for the quantum geometry of the horizon we use the quantum version of (\ref{curvaturewithpunctures})
\be\label{qIHBC} 
\bigl(\hat{F}^i+\frac{4\pi}{k}\sum_{p\in\mathcal{P}}\delta^2(x,x_p)\hat{J}^i_{\rho_{p}}\bigr)\psi_{\rm horizon}=0,
\ee
which selects elements of the physical Hilbert space of the horizon theory. Notice that at each puncture $p$ the angular momentum algebra $[\hat{J}^i_{\rho_{p}},\hat{J}^j_{\rho_{p}}]=\epsilon^{ij}_k\hat{J}^k_{\rho_{p}}$ holds. The quantum version of (\ref{IHBC})
\be\label{IHBConstates}
\biggl(\mathbb{I}\otimes\widehat{F}^i+\frac{\pi(1-\gamma^2)}{a_H}\widehat{\Sigma}^i\otimes \mathbb{I}\biggr)\psi_{\rm total}=0,
\ee
with $\psi_{\rm total}=\psi_{\rm bulk}\otimes\psi_{\rm horizon}$ couples bulk and horizon quantum d.o.f. properly back together.\footnote{In fact, only the exponentiated version of $\widehat{F}^i$ is well-defined \cite{ABK} but the subsequent discussion will not be altered by this.} The physical Hilbert space is then given by 
\be\label{physHS}
\mathcal{H}_{\rm phys}=\biggl(\bigoplus_\mathcal{P}\mathcal{H}^{\mathcal{P}}_{\rm bulk}\otimes\mathcal{H}^{\mathcal{P}}_{\rm horizon}\biggr)/\mathcal{G}_{\rm total},
\ee
where $\mathcal{H}^{\mathcal{P}}_{\rm bulk}$ denotes the bulk space of states. One denotes by $\mathcal{G}_{\rm total}=\mathcal{G}_{\rm bulk}\ltimes\mathcal{G}_{\rm horizon}$ internal $SU(2)$-transformations, diffeomorphisms which preserve the surface and eventually motions, generated by the Hamiltonian constraint $H$ \cite{ABK,SU2}. Since the IH framework stipulates that the lapse is restricted to vanish on the horizon, the scalar constraint $H$ is only imposed in the bulk. In fact, the horizon states satisfying the boundary condition (\ref{qIHBC}) are automatically gauge and diffeomorphism invariant since these invariances on the horizon are implemented by the same operators (\ref{GaussConstraint},\ref{VectorConstraint}) as (\ref{qIHBC}). After imposition of the respective constraints one has
\be
\mathcal{H}_{\rm phys}=\bigoplus_N\bigoplus_{(j)_1^N}\mathcal{H}_{\rm bulk,phys}^{(j)_1^N}\otimes {\rm Inv}_k(\otimes_p j_p),
\ee
where ${\rm Inv}_k(\otimes_p j_p)$ is the CS-Hilbert space on the punctured sphere with $j_p\leq k/2$ and $\mathcal{H}_{\rm bulk,phys}^{(j)_1^N}$ denotes the physical Hilbert space of the bulk for a corresponding puncture configuration \cite{PranzettiPolo,Commentoncoupling}.\newline
In order to analyze the thermodynamical properties of the horizon one computes the total number of (micro-) states available to it given by
\be\label{dimension}
W(\{\mathcal{P}\})=\sum_{\mathcal{P}}\dim({\rm Inv}_k(\otimes_p j_p),
\ee
where we constrain ourselves only to those horizon states which are compatible with $a_H$ and $j_p\leq k/2$. In the following let $n_j$ denote the occupation number of a certain puncture type that is labeled by an irreducible representation $\rho_j$ of $su(2)$. Then (\ref{dimension}) can be rewritten for a quantum configuration $\{n_j\}$ as
\be\label{AnzahlMicrozustaende}
W({\{n_{j}\}})=\frac{N!}{\prod_j n_j!}~\frac{2}{k+2} \sum_{l=0}^{k/2}\sin^2\biggl(\frac{(2l+1)\pi}{k+2}\biggr)\prod_{j} d^{n_j}_j(l),
\ee
wherein
\be
d_j(l)\equiv\biggl[\frac{\sin(\frac{(2j+1)(2l+1)\pi}{k+2})}{\sin(\frac{(2l+1)\pi}{k+2})}\biggr]
\ee
as in \cite{KaulMajumdar,SU2,SU2rev}. The total number of punctures is denoted by $N=\sum_j^{k/2} n_j$ and the combinatorial pre-factor indicates that in the purely gravitational case the punctures are considered as distinguishable \cite{Rovelli,ABK,KrasnovAPS, Pithis}. Since the level of $k$ the theory is proportional to $a_H/\ell_p^2$ it is convenient to assume the limit $k\to\infty$ of $(\ref{AnzahlMicrozustaende})$ giving
\be\label{AnzahlMicrozustaendekgross}
W({\{n_{j}\}})=N!\prod_j\frac{(2j+1)^{n_j}}{n_j!},
\ee
where we neglected the next-to-leading order term in $k$ which would give rise to the notorious logarithmic correction of the entropy \cite{SU2, KaulMajumdar, logcorrection, Mitra}.
It counts the number of distinct microstates belonging to the distribution set $\{n_j\}$ and is that of a typical Maxwell-Boltzmann statistics for distinguishable entities \cite{StatMech}.\newline
The introduction of proper notions of a quasi-local energy and a local temperature
\be
E=\frac{A}{8\pi\ell},~~T=\frac{\ell_p^2}{2\pi\ell} 
\ee
of the isolated horizon associated with a stationary observer at distance $\ell$ from the horizon facilitates its statistical mechanical analysis \cite{APE, espera, Pranzetti}. With the horizon area spectrum
\be
\widehat{A}|\{n_j\}\rangle=8\pi\ell_p^2\gamma\sum_j n_j \sqrt{j(j+1)}|\{n_j\}\rangle
\ee
the canonical partition function reads
\be
Z(T,N)=\sum_{\{n_j\}}^{-}W(\{n_j\})e^{-\beta E},
\ee
where the primed summation goes over all distribution sets that conform to the restrictive condition $N=\sum_j n_j$. With this the expression for the entropy is obtained as
\be
S=-\beta^2\partial_{\beta}(\log Z/\beta)=A/(4\ell_p^2)+\sigma(\gamma)N,
\ee
where $\sigma(\gamma)=\log[\sum_j (2j+1) e^{-2\pi\gamma\sqrt{j(j+1)}}]$. This result is in agreement with the corresponding results using the microcanonical and grandcanonical ensemble. Furthermore, the entropy function is both extensive in $A$ and $N$ as it should in order to agree with the laws of (phenomenological) thermodynamics and black hole mechanics. Remarkably, it is compatible with the semiclassical Bekenstein-Hawking area law \cite{Bekenstein, Hawking}, since the second summand only expresses the quantum hair correction due to the quantum geometry of the isolated horizon.

\subsection{Distinguishability of horizon states}\label{subsectionC}
We have tacitly used until here that the horizon states are distinguishable in the purely gravitational case. This choice drastically influences the quantum statistics of the model and is well motivated in the LQG literature \cite{ABK,Rovelli,KrasnovAPS,Commentoncoupling,Pithis}. We want to revise the supporting arguments here.\newline
Firstly, as far as naive state counting is concerned, only under the assumption of distinguishable states does one obtain a linear entropy/area-relation \cite{KrasnovAPS} and an extensive entropy function \cite{Pithis}.\newline
Secondly, the distinguishability of the horizon states follows from a purely algebraic point of view through the boundary condition \cite{Rovelli,Commentoncoupling}. To see this, take (\ref{physHS}) and think of $A$ as $\mathcal{H}_{{\rm bulk}}^{\mathcal{P}}$, $B$ as $\mathcal{H}_{{\rm horizon}}^{\mathcal{P}}$, $\times$ as $\otimes$ and $G$ as $\mathcal{G}_{{\rm total}}$, whose action we assume not to be free at first. Then we have the following isomorphism 
\be\label{ABGnonfree}
(A\times B)/G \cong (A/G)\times \biggl(\bigcup_{[a]\in A/G}B/G_{a}\biggr) ,
\ee
where $G_{a}$ is the stabilizer of $a\in A$. If the action of $G$ is free, all stabilizers are trivial and (\ref{ABGnonfree}) simplifies to
\be
(A\times B)/G \cong (A/G)\times B.
\ee
Applying this now to the underlying algebraic structure of the isolated horizon state space, we start with (\ref{ABGnonfree}) and think of the elements in $A,B$ again as states. Then $(a,b)$ and $(a',b')$ are only in the same $G$-orbit, if one requires that the surface state $b'\in \mathcal{H}^{\mathcal{P}}_{\rm horizon}/G_a$. Any $(a'',b'')$ with $b''\notin \mathcal{H}^{\mathcal{P}}_{\rm horizon}/G_a$ does not lie in the same $G$-orbit as $(a,b)$ and is distinct from the latter. Consequently, physically equivalent states, i.e. states in the same $G$-orbit, can be formally discriminated from states in different orbits. Going back to our case, one has 
\be
(\mathcal{H}^{\mathcal{P}}_{\rm bulk}\otimes \mathcal{H}^{\mathcal{P}}_{\rm horizon})/\mathcal{G}_{\rm total} \cong(\mathcal{H}^{\mathcal{P}}_{\rm bulk}/\mathcal{G}_{\rm bulk})\otimes {\rm Inv}_k(\otimes_p j_p),
\ee
where ${\rm Inv}_k(\otimes_p j_p)\cong\mathcal{H}^{\mathcal{P}}_{\rm horizon}/\mathcal{G}_{\rm horizon}$ is just the CS-Hilbert space. Gauge and (spatial) diffeomorphism constraints in LQG generate transformations, which are connected to the identity. Therefore $\mathcal{G}_{\rm total}$'s action is free and only trivial stabilizers acting on the horizon states leave the bulk invariant.\newline
As a consequence, a specific transformation can transform a state $b$ of the horizon Hilbert space in one orbit into a different one in another orbit. Such a transformation could be a permutation of the puncture labels which generally does not leave the states invariant. The observable $\widehat{F}^i$ acting on them does not commute with a permutation unless two equivalent puncture labels are interchanged or the permutation is trivial. This is due to the fact, that in the specific case of the horizon punctures one deals with identical but nonetheless distinguishable entities \cite{Rovelli,ABK,KrasnovAPS,Pithis}. This means that they do not obey the \emph{indistinguishability postulate} for identical quantum objects \cite{StatMech,QstatTaxonomy}. Identical quantum objects satisfy the former postulate, if \emph{all} observables $\mathcal{O}$ commute with \emph{all} permutations $P$ of the considered entities, i.e. $[\mathcal{O},P]=0$. Then observables are just symmetric Hermitian operators. In our model the relevant observables of the horizon model for the discussion of its statistics are the operators for the area $\widehat{A}$ and the field strength $\widehat{F}^i$, respectively. It is clear that $[\widehat{A},P]=0$ holds but for $\widehat{F}^i$ this would be false in general, hence distinguishability despite identicality.\newline
To see this, assume we knew the wave functional $\psi_0$ describing the case without sources associated to the state $|0\rangle$. Then eq. (\ref{curvaturewithpunctures}) implies $\widehat{F}^i\psi_{0}=0$. However, in the presence of point-like sources at
$\{x_p\}_{1}^N$, $\widehat{G}^i$ acts on the horizon wave functional $\psi_{\rm horizon}$ as
\begin{align}\label{Gactiononpsi}
&\widehat{G}^i\psi_{\rm horizon}=\\ &\biggl(\widehat{F}^i+\frac{4\pi}{k} (\delta^2(x,x_1)\hat{J}^i_{\rho_{1}}\otimes\mathbb{I}_2\otimes...\otimes\mathbb{I}_N+...)\biggr)\psi_{\rm horizon}=0.\nonumber
\end{align}
Consider now the Wilson line operator $h_{\gamma_p}(A_{\rho_p})=Pe^{i\int_{\gamma_{p}}A_{\rho_p}}$, where $A_{\rho_p}=A_i\widehat{J}^i_{\rho_p}$, $\gamma_p$ is a non-intersecting path on the punctured $S^2$ and its end point marks the position of the $p$th source. $P$ denotes the path-ordered product. Applying the gauge transformation prescription
\be
[\widehat{G}^i,h_{\gamma_p}(A_{\rho_p})]=h_{\gamma_p}(A_{\rho_p})\hat{J}^i_{\rho_{p}}\delta^2(x,x_p)
\ee
to it, one constructs the following product wave functional 
\begin{align}\label{productstate}
\psi_{\rm horizon}=\psi_{1}(A_{\rho_{1}})\otimes...\otimes\psi_{N}(A_{\rho_{N}})~\psi_{0},
\end{align}
wherein
\be
\psi_p(A_{\rho_{p}})=\langle A_{\rho_{p}}|\psi_p\rangle \equiv h_{\gamma_p}(A_{\rho_p})
\ee
holds. Hence, it corresponds to the $N$-puncture state $|\{p,\rho_p\}_{p=1...N}\rangle$. Equipped with this and when ignoring the pre-factor, $\hat{F}^i$ acts on $\psi_{\rm horizon}$ as
\be
\hat{F}^i\psi_{\rm horizon}=(\delta^2(x,x_1)\hat{J}^i_{\rho_{1}}\otimes\mathbb{I}_2\otimes\ldots\otimes\mathbb{I}_N+\ldots)\psi_{\rm horizon}.
\ee
\noindent
To show that the punctures are actually distinguishable, consider for simplicity the case $N=2$ and further let $P_{12}$ swap the arguments of the first and second puncture. This yields
\begin{align}
&P_{12}\hat{F}^i\psi_{\rm horizon}=\\
&(\delta^2(x,x_1)\hat{J}^i_{\rho_{1}}\otimes\mathbb{I}_2+\mathbb{I}_1\otimes \hat{J}^i_{\rho_{2}}\delta^2(x,x_2))\psi_1(A_{\rho_{2}})\otimes\psi_2(A_{\rho_{1}})\nonumber,
\end{align}
whereas
\begin{align}
&\hat{F}^i P_{12}\psi_{\rm horizon}=\\&(\delta^2(x,x_1)\hat{J}^i_{\rho_{2}}\otimes\mathbb{I}_2+\mathbb{I}_1\otimes \hat{J}^i_{\rho_{1}}\delta^2(x,x_2))\psi_1(A_{\rho_{2}})\otimes\psi_2(A_{\rho_{1}})\nonumber.
\end{align}
Consequently, for generic permutations and horizon wave functions we have the simple yet important result
\be\label{indiscond} 
[\widehat{F}^i,P_{pp'}]\psi_{\rm horizon}\neq 0,
\ee
unless $\hat{J}^i_{\rho_{p}}=\hat{J}^i_{\rho_{p'}}$. Hence, $\psi_{\rm horizon}$ and $\widehat{F}^i$ are non-symmetric with respect to distinct representations.\footnote{For indistinguishable punctures, the boundary condition and states would have to get totally (anti-)symmetrized. Considering just gravitational d.o.f. the state counting then leads to a non-linear entropy/area-relation \cite{KrasnovAPS}.} When keeping the representations attached to the incident bulk links (supporting $\widehat{\Sigma}^i$) locked, but arbitrarily permuting horizon puncture labels, the boundary condition (\ref{IHBConstates}) would in some cases get violated. In other words, bulk and surface d.o.f. cannot get arbitrarily coupled \cite{ABK,SU2,Commentoncoupling}. This imposes an ordering relation onto the set of punctures which we denote here by $\widehat{N}$. Small diffeomorphisms are elements of ${\rm Diff}_0(S^2_{\widehat{N}})$ and cannot change the order of the puncture set but a non-trivial permutation can do this. It follows, that a microstate (which is a representative in one $G$-orbit) is changed by such a puncture permutation into a physically distinct (non-diffeomorphic) one lying in a different orbit. From the statistical point of view one actually has to count such different equivalence classes which correspond to different microstates and that are accessible to the system in the macrostate $(E,N)$. This is reflected by the statistical distributions (\ref{AnzahlMicrozustaende}) and (\ref{AnzahlMicrozustaendekgross}), respectively.\newline
Notice however, that only in $d\geq 3$ spatial dimensions particle exchange is mediated by an element $P$ of the permutation group $S_N$. In the next section we account for the fact, that in $d=2$ such an exchange is mediated by (the generalization of $S_N$ to) the braid group. Motivated by the previous discussion, we will still assume that differently labeled punctures are distinguishable.

\section{Anyonic statistics and LQG black holes}\label{SectionIII}
\subsection{Appearance of anyonic statistics}\label{subsectionA2}
\begin{figure}[htbp]
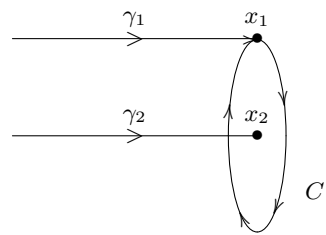

\centering
\[
\xy
/r1.2pc/:{\xypolygon4"B"{~:{(4.5,0):(0,.4)::}~>{}}},
"B1";"B2"**@{-},"B3";"B4"**@{-},
"B4"*{\bullet}*\xycircle(.75,2.52){+++++++\dir{<}}*\xycircle(.75,2.52){-},
"B1"*{\bullet},
"B1";"B2"**@{}?<>(.5)="C1"*{>},"B3";"B4"**@{}?<>(.5)="C2"*{>},
"B2"+(0,.5)*{},"B3"+(0,.5)*{},
"C1"+(0,.5)*{\gamma_1},"C2"+(0,.5)*{\gamma_2},
"B1"+(0,.5)*{x_1},"B4"+(0,.5)*{x_2},"B1"+(1.5,-4)*{C}
\endxy 
\]
\caption{Parallel transport of the first puncture around the second one.}
\label{fig:monodromy}
\end{figure}

The dimensionality of the problem and the observation that the (topological) source term in (\ref{ActionPunctures}) carries $\hbar$ explicitly in it, steer us into a closer investigation of the features associated with the topology of the phase space because it reveals the inherent possibility of having anyonic statistics for the horizon d.o.f..\newline
We will adopt the stance that quantum statistics refers to the phase, which arises when two particles of a multi-particle quantum system are exchanged with each other. Hence, this section will be concerned with explaining how this phase arises in our system and how it can be computed through the \textit{Knizhnik-Zamolodchikov} connection, which leads to a notion of parallel transport and thus puncture exchange. The key idea is that due to the topological defects/punctures, connections become elements of the non-trivial first de Rham cohomology group on the phase space (\ref{phasespacepuncturessm}). Then we connect the former group to the fundamental group of the configuration space whose representations label in the quantum theory inequivalent quantizations. We give a prescription to compute these representations and link them to anyonic statistics. To see this clearly, we have to analyze at first features of phase spaces with topological defects like (\ref{phasespacepuncturessm}) and therefore we import tools from symplectic geometry \cite{sympgeom} now.\newline
To this aim, consider a generic symplectic manifold $(\Gamma,\omega)$. One calls a vector field $\eta$ on $\Gamma$ that preserves $\omega$, i.e. $L_{\eta}\omega=0$, a symplectic vector field. Using Cartan's magic formula and the closedness of $\omega$ one has
\be
L_{\eta}\omega=d(i_{\eta}\omega)=0.
\ee
$\eta$ is only symplectic if $i_{\eta}\omega$ is closed, whereas it is a Hamiltonian vector field, if additionally $i_{\eta}\omega$ is exact. It is a fact, that locally on every contractible (i.e. simply connected) open set, symplectic vector fields are Hamiltonian. Additionally, if one has trivial first de Rham cohomology group, i.e $H^1(\Gamma; \mathbb{R})=0$, then globally every symplectic vector field is Hamiltonian and we can write $i_{\eta}\omega=-d f$, for some function $f\in C^{\infty}(\Gamma,\mathbb{R})$. The diffeomorphisms of $\Gamma$, which are generated by Hamiltonian vector fields are known as canonical transformations. However, in the case that $H^1(\Gamma; \mathbb{R})\neq 0$, for some transformations $\eta$ the corresponding $i_{\eta}\omega$ is a non-trivial element of $H^1(\Gamma; \mathbb{R})$ and therefore there is no globally defined function $f$ on $\Gamma$ for this transformation. Equivalently, there can be several choices for the canonical $1$-form $\theta$ differing by elements of $H^1(\Gamma;\mathbb{R})$, but giving rise to the same symplectic $2$-form $\omega$. The ambiguity in $\theta$ has no effect on the classical equations of motions but nevertheless $H^1(\Gamma; \mathbb{R})$ 'measures' the obstruction for symplectic vector fields to be Hamiltonian.\newline 
Let us apply this to the one-form (\ref{CSsymp1form}) on the phase space without defects (\ref{phasespacewithoutpunctures}). If $\tilde{\theta}$ is closed, then $\theta$ and $\theta+\tilde{\theta}$ will lead to the same $\omega$. If $\tilde{\theta}$ was closed and exact, we could figure $\tilde{\theta}$ as $\tilde{\theta}=\delta\rho[A]$, where $\rho[A]$ is some globally defined function(al) on (\ref{phasespacewithoutpunctures}) and the connection $A$ lives on $M\cong\mathbb{R}\times \Sigma$. The function $\rho[A]$ is a canonical transformation and one can transform $\delta\rho[A]$ to $0$, as implied by Poincar\'e's lemma. \textit{On the contrary}, in the case of the isolated horizon we have $\Delta\cong\mathbb{R}\times S^2$ with punctures (i.e. topological defects) on it, so we have to consider (\ref{phasespacepuncturessm}). There $\tilde{\theta}$ is closed but due to the defects not exact. Therefore, $\tilde{\theta}$ is a non-trivial element of the de Rham cohomology $H^1(\Gamma;\mathbb{R})$ and it cannot be transformed to $\tilde{\theta}=0$ upon canonical transformation. One can only locally write $\tilde{\theta}=\delta\rho[A]$, since $\rho[A]$ is not globally definable. This is of relevance for the quantum theory of generic anyonic systems \cite{Nair} and also for our problem, as we see below, because the defects lead to non-trivial $H^1(\Gamma; \mathbb{R})$ and thus non-contractible loops on $\Gamma$.\newline
To see this, the phase space (\ref{phasespacepuncturessm}) is reparametrized by holonomies \cite{AlekseevSchomerus,AGN} which yields
\begin{align}\label{phasespacepuncturestop}
&\Gamma=\\
&\{\rho\in {\rm Hom}(\pi_1(\mathcal{F}_N(S^2)),SU(2))|\rho(c_p)\in\mathcal{C}_p^G\}/SU(2).\notag
\end{align}
The $\{c_p\}$ stand for the generators of ${\rm Hom}(\pi_1(\mathcal{F}_N(S^2)),SU(2))$ which concur with such non-contractible oriented loops around the punctures $\{p_i\}_1^N$. $\mathcal{F}_N(S^2)$ denotes the configuration space (cf. Appendix (\ref{AppendixC})). For the specific case of distinguishable puncture species $\{n_j\}_{1/2}^{k/2}$ distributed on $S^2$ it reads as
\be\label{configspacepuncturedspehre}
\mathcal{F}_N(S^2)=\{(x_1,...,x_N)\in (S^2)^N|x_p\neq x_{p'}~for~p\neq p'\}.
\ee 
The fundamental group $\pi_1$ of this space is the spherical braid group 
\be\label{SBGMS}
B_{n_{1/2},...,n_{k/2}}(S^2)
\ee 
on $N$ strands (cf. Appendixes (\ref{AppendixB},\ref{AppendixD})).\newline
If we invoke the perspective of topology \cite{MathPhys,AnyonsI,AnyonsII} for the quantization of our problem as reviewed in (\ref{subsectionB}), a kinematical ambiguity in the quantization of the classical system on the configuration space $\mathcal{F}_N(S^2)$ arises and has to be classified by the set of all irreducible unitary representations of the fundamental group $\pi_1(\mathcal{F}_N(S^2))$. According to this, the respective quantum theory deals with multi-component state vectors lying in $SU(2)$ and these are labeled by a non-abelian phase $\rho\in {\rm Hom}(\pi_1(\mathcal{F}_N(S^2)),SU(2))$, with $\pi_1(\mathcal{F}_N(S^2))\cong B_{n_{1/2},...,n_{k/2}}(S^2)$. It is this phase which accounts for the non-abelian anyonic statistics of the horizon puncture system \cite{AnyonsI, AnyonsII, MooreRead, NayakTQC}.\footnote{If $A$ was a $U(1)$-connection, the phase would be abelian and it is well established that such a phase is needed to describe particles of abelian anyonic statistics in $2d$ \cite{AnyonsI}.} We want to emphasize, that for this classification \emph{no knowledge} of the dynamics of the system is needed. If the black hole was still modeled by a topological $2$-sphere with punctures but different constraints, such phases would still show up. The account of $\Gamma$'s topological intricacies thus solely unveils the anyonic nature of the LQG horizon degrees of freedom.\newline 
In the following, we will discuss how to compute such non-abelian phases, which actually correspond to the parallel transport of punctures along and around each other. To illustrate this, consider the winding of one puncture along a loop $C$ completely around a second one as in Fig. (\ref{fig:monodromy}), where other punctures are suppressed.
On the quantum level this transformation is translated as 
\be
h_{C\circ\gamma_1}(A_{\rho_{1}})h_{\gamma_2}(A_{\rho_{2}})\prod_{p>2}h_{\gamma_p}(A_{\rho_{p}})~\psi_0.
\ee
Commuting $h_C(A_{\rho_{1}})$ with $h_{\gamma_2}(A_{\rho_{2}})$ would lead to a non-abelian phase \cite{Guadagnini, Broda, NayakTQC}. Equivalently, we can consider the configuration space (\ref{configspacepuncturedspehre}) and rewrite it by means of $S^2\cong\mathbb{C}\cup\{\infty\}$ as
\be
\mathcal{F}_N(S^2)\cong\{(z_1,...,z_N)\in (S^2)^N|z_p\neq z_{p'}~for~p\neq p'\}.
\ee
This is equivalent to 
\be
\{(S^2)^N-\bigcup_{1\leq p<p'\leq N}K_{pp'}\},
\ee
where $K_{pp'}=\{(z_1,...,z_N)\in (S^2)^N|z_p=z_{p'}\}$. The form of the phase space (\ref{phasespacepuncturestop}), allows us to trade $H^1(\Gamma,\mathbb{R})$ for $H^1(\mathcal{F}_N(S^2),\mathbb{R})$. The closed holomorphic $1$-form
\be
\omega_{pp'}=\frac{1}{2\pi i}d~\log(z_p-z_{p'})
\ee
on $\mathcal{F}_N(S^2)$ represents the de Rham cohomology class of generators $\omega_{pp'}\in H^1(\mathcal{F}_N(S^2);\mathbb{Z})$ with $1\leq p<p'\leq N$ \cite{Arnold}. We thus introduce the Knizhnik-Zamolodchikov or Kohno connection to the context of LQG black holes as
\be\label{KZKconnection}
\widehat{A}_K=\frac{4\pi}{k+2}\sum_{1\leq p<p'\leq N}\hat{J}^i_{\rho_{p}}\otimes \hat{J}^i_{\rho_{p'}}~\omega_{pp'},
\ee
wherein $\otimes$ denotes the Kronecker product \cite{CFT, Guadagnini}.\newline
Using this, a simultaneous puncture rearrangement can be given using the holonomy operator of (\ref{KZKconnection})
\be\label{KZKholonomy}
\hat{\rho}(A_K,\gamma)=P~e^{i\oint_{\gamma}{A}_K},
\ee
where the loop $\gamma$ is taken from the homotopy class $[\gamma]\in B_{n_{1/2},...,n_{k/2}}(S^2)$.\footnote{Notice however, that the contour integral of a meromorphic $1$-from such as $\omega_{pp'}$ on a compact surface $S^2$ along a loop $\gamma$ that encircles all poles vanishes. The sum over all residues yields $0$ because such a loop can always be shrunk to a point on the back of the sphere (cf. Appendix (\ref{AppendixD})).} Hence, by exchanging/moving the horizon d.o.f. on the punctured $2$-sphere, the wave function picks up a non-abelian phase, namely
\be
\psi_{\rm horizon}\to\hat{\rho}_{[\gamma]}(A_K)\psi_{\rm horizon},
\ee
which specifies the statement made above that inequivalent quantizations on the multiply connected configuration space $\mathcal{F}_N(S^2)$ are marked by representations $\{\rho:B_{n_1,...,n_{k/2}}(S^2)\to SU(2)\}$. We have thus clarified, how (non-abelian) anyonic statistics is encoded in the description of the LQG black hole model based on a puncture system representing the quantum d.o.f. of the horizon. We want to further investigate this exchange behavior now by relating it to the large diffeomorphisms of the punctured horizon.

\subsection{Large diffeomorphisms and the braid group}\label{subsectionB2}
\begin{figure}[htbp]
\centering
\includegraphics[clip,width=0.45\textwidth]{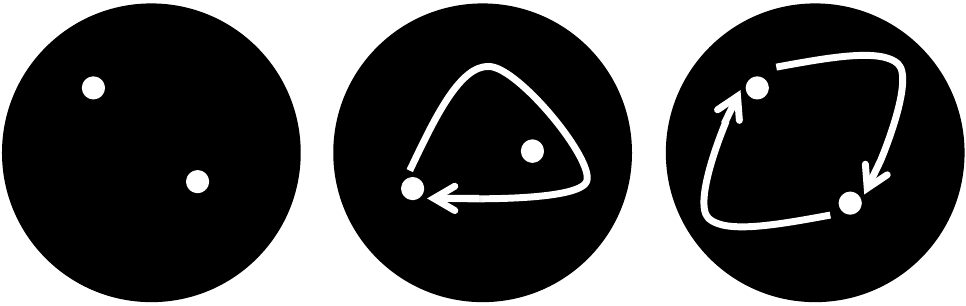}
\caption{Two horizon punctures: unbraided vs. upon the application of $\widehat{M}$ and $\widehat{B}$, respectively.}
\label{punctureddisc1}
\end{figure}

\begin{figure}[htbp]
\centering
\includegraphics[clip,width=0.45\textwidth]{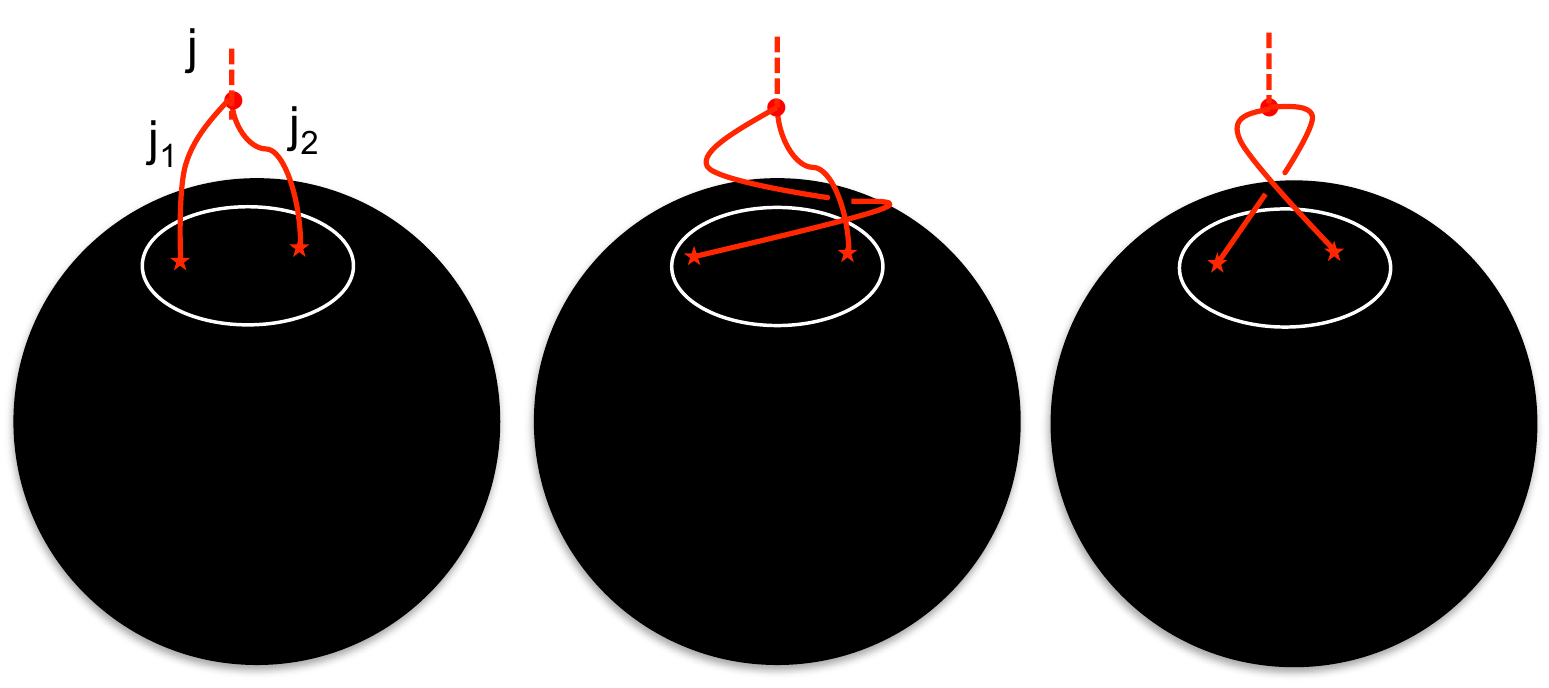}
\caption{Two incident bulk edges piercing the horizon: unbraided vs. upon the application of $\widehat{M}$ and $\widehat{B}$, respectively.}
\label{punctureddisc2}
\end{figure}

Motivated by our preliminary discussion of the symmetries of CS-theory and the fact that the horizon puncture system is invariant with respect to small diffeomorphisms, we want to take a closer look onto the action of the large diffeomorphisms on our system. These sorts of diffeomorphisms of the punctured $2$-sphere fall into the mapping class group $M_{n_{1/2},...,n_{{k/2}}}(S^2)$, which we discuss in Appendix (\ref{AppendixE}). A priori, horizon states could either be invariant under it or transform by a unitary representation of it \cite{Smolin}. In the former case, large diffeomorphisms would be considered as gauge, whereas in the latter they would be regarded as a symmetry of the theory for which we will argue below.\newline
The action principle does not dictate the transformation properties of the physical states under the diffeomorphisms which are not in the identity component. This is because no constraints are associated to them. Small diffeomorphisms are generated by the constraints encoded in the action (\ref{VectorConstraint}), so only they should a priori be factored out. To demand the invariance under large diffeomorphism transformations would amount to an extra assumption \cite{TeitelboimHenneaux}. On the classical level a diffeomorphism of the punctured $S^2$ induces a linear transformation on $H^1(\mathcal{F}_N(S^2),\mathbb{R})$, which in turn is the reason why the latter gives rise to a representation of the mapping class group \cite{Witten}.\newline
Interestingly, the discussion of these specific diffeomorphisms can be easily connected to the previous discussion of the statistical symmetry of the puncture system, which is given by its braid group. In the Appendix we discuss and explicitly recover in (\ref{finalmappingclassgroup}) that these groups are related as
\be\label{MCGBGS}
M_{n_{1/2},...,n_{{k/2}}}(S^2)\cong B_{n_{1/2},...,n_{{k/2}}}(S^2)/\mathbb{Z}_2.
\ee
Let us exemplify this point by considering $N$ punctures on one hemisphere of $S^2$. This would be homeomorphic to a $N$-punctured disc. From algebraic topology one knows, that the mapping class group of the $N$-punctured disc $M_N(D^2)$ is isomorphic to the braid group of the disc $B_N(D^2)$ on $N$ strands which in turn is equivalent to $B_N(\mathbb{R}^2)$. Hence, for this topology the statistical symmetry of the puncture system is given by the large diffeomorphisms. Using the tools given in the last subsection, we are able to calculate unitary representations of braiding generators e.g. for the setting of $2$ colored punctures. By executing the contour integral in (\ref{KZKholonomy}) in the case of two punctures, one yields the monodromy operator
\be\label{MonodromyMatrix}
\widehat{M}_{(1,2)}\psi\equiv\hat{\rho}(A_K,\sigma_1^2)\psi=q^{2~\hat{J}^i_{\rho_{1}}\otimes \hat{J}^i_{\rho_{2}}}\psi,
\ee
where $\sigma_1$ is a generator of the braid group (cf. Appendix (\ref{AppendixB})), $q=e^{i\frac{2\pi}{k+2}}$ is the so-called deformation parameter and we dropped the subscript of the wave function.\footnote{If we consider e.g. the case where both punctures are colored with the fundamental representation of su(2), we obtain for the monodromy with (\ref{KZKconnection},\ref{KZKholonomy})
\be
\widehat{M}=
\begin{pmatrix}
  q^{1/2} & 0 & 0 & 0 \\
  0 & \frac{1}{2}(q^{1/2}+q^{3/2}) & \frac{1}{2}(q^{1/2}-q^{-3/2}) & 0 \\
  0  & \frac{1}{2}(q^{1/2}-q^{-3/2})  & \frac{1}{2}(q^{1/2}+q^{3/2}) & 0  \\
  0 & 0 & 0 & q^{1/2}
 \end{pmatrix}
 \ee
with eigenvalues $q^{1/2}$ (triplet) and $q^{-3/2}$ (singlet).}\footnote{Higher powers of $\widehat{M}$ correspond to different winding numbers and encirclement of several punctures corresponds to the ordered product of monodromy operators. That these representations can generically be well-defined is guaranteed by (\ref{wohldefdarst}).} Since $\widehat{M}$ represents the case of two consecutive exchanges of puncture $1$ with $2$, the braiding matrix is
\be\label{BraidingMatrix}
\widehat{B}\equiv\hat{\rho}(A_K,\sigma_1)=q^{{\hat{J}^i_{\rho_{1}}\otimes \hat{J}^i_{\rho_{2}}}}~P_{12},
\ee
where $P_{12}$ is the permutation operator. We depict the effect of $\widehat{M}$ and $\widehat{B}$ as in Fig. (\ref{punctureddisc1}).
In the limit of large black holes, i.e. $k\to\infty$, the operator $\widehat{M}$ is just the identity. Since $\widehat{B}$ is also affected in this limit, a puncture exchange will be solely mediated by a non-abelian representation of the permutation operator $P_{12}$. Pictorially, in the case of large black holes the topological information about what happened along the braid is forgotten. Braids with the same initial and final configurations but different windings are identified and the same applies to the corresponding mapping classes, too. Hence, we infer that the braiding is a \textit{quantum effect} becoming relevant for small (and smaller becoming) black holes and that the information about their state is thus not solely of combinatorial nature.\newline
Assume an appropriately defined (physical) inner product as in \cite{CSIP, AlekseevSchomerus} such that
\begin{align}
\langle \psi_1|\psi_2 \rangle &=\prod_{p=1}^N\int\limits_{SU(2)}d\mu(h_{\gamma_p})\overline{D^{j_{p}'}_{m' n'}(h_{\gamma_{p}}(A))}D^{j_{p}}_{m n}(h_{\gamma_{p}}(A)),
\end{align}
where the Wigner matrices $D^{j_{p}}_{m n}(h_{\gamma_{i}}(A))$ give the spin-$j_{p}$ irreducible matrix representation of the $SU(2)$ group element $h_{\gamma_{p}}(A)$. Due to Schur's orthogonality relation this expression vanishes if the respective operators $h_{\gamma_p}$ in $\psi_1$ and $\psi_2$ do not carry the same representation. Hence, the mapping class group separates the relevant Hilbert space into orthogonal subspaces via $\widehat{B}$-type operators as
\be
\langle \psi|\widehat{B}\psi\rangle=0~~\text{if}~~ \hat{J}_{\rho_{p}}\neq\hat{J}_{\rho_{p'}}.
\ee
Since the action of large diffeos on punctures will also drag the incident bulk edges along, this will cause a knotting of the spin network at least in the vicinity of the horizon as in Fig. (\ref{punctureddisc2}), which cannot be unraveled through a (small) bulk diffeomorphism.
In the following we want to investigate whether such a different knotting of the spin network in the neighborhood of the horizon has any observable consequences. The area operator would not be of great help here, since $\widehat{A}$ is a function of the $su(2)$-Casimir operator and thus commutes with all the generators of this Lie algebra. For a representation $\hat{\rho}$ of a generic element of the braid group one has 
\be\label{mcgtrafoA}
\langle \hat{\rho}\psi | \widehat{A}|\hat{\rho}\psi\rangle=\langle\psi|\hat{\rho}^{-1}\widehat{A}\hat{\rho}|\psi\rangle=\langle \psi |\widehat{A}|\psi\rangle.
\ee
The area operator $\widehat{A}$ is thus also invariant under large diffeomorphisms. In contrast to this, we will see that the field strength $\widehat{F}^i$ is an observable which could in principle be used to distinguish between different knottings since it transforms under the action of large diffeomorphisms/braid group of the punctured surface as
\be
\widehat{F}^i\to\widehat{F'}^i=\hat{\rho}^{-1}\widehat{F}^{i}\hat{\rho}.
\ee
To this aim, we take the difference of the expectation values
\be
\langle\hat{\rho}^n\psi|\widehat{F}^i|\hat{\rho}^n\psi\rangle-\langle\psi|\widehat{F}^i|\psi\rangle,
\ee
where $\hat{\rho}^n=\bigl(q^{{\widehat{J}^j_{\rho_{1}}\otimes \widehat{J}^j_{\rho_{2}}}}P_{pp'}\bigr)^n$ with $n=1,2$. Without loss of generality we set $N=2$ and when using the Baker-Campbell-Hausdorff formula and its Hadamard lemma one yields
\be
\langle\psi|\sum_{m=0}^{\infty}\frac{(\frac{n}{i}\frac{2\pi}{k+2})^m}{m!}[\widehat{J}^j_{\rho_{1}}\otimes \widehat{J}^j_{\rho_{2}},(P_{12}^{-1})^{n}\widehat{F}^i(P_{12})^{n}]_m-\widehat{F}^i|\psi\rangle
\ee
where $[\widehat{J}^j_{\rho_{1}}\otimes \widehat{J}^j_{\rho_{2}},(P_{12}^{-1})^{n}\widehat{F}^i(P_{12})^{n}]_0\equiv(P_{12}^{-1})^{n}\widehat{F}^i(P_{12})^{n}$ and $[\widehat{J}^j_{\rho_{1}}\otimes \widehat{J}^j_{\rho_{2}},(P_{12}^{-1})^{n}\widehat{F}^i(P_{12})^{n}]_m\equiv[\widehat{J}^j_{\rho_{1}}\otimes \widehat{J}^j_{\rho_{2}},[\widehat{J}^j_{\rho_{1}}\otimes \widehat{J}^j_{\rho_{2}},(P_{12}^{-1})^{n}\widehat{F}^i(P_{12})^{n}]_{m-1}]$.
Multiplying from the left with $\hat{\rho}^n$ gives
\be\label{indiscond2}
[\widehat{F}^i,\hat{\rho}^n]\neq 0,
\ee
unless $n=2$ and $k\to\infty$. For $n=1$ and $k\to\infty$ (\ref{indiscond2}) reduces to expression (\ref{indiscond}). For example, when $n=2$ the commutator yields
\begin{align}
&[\widehat{F}^i,\widehat{M}]=i\frac{4\pi}{k+2}\frac{4\pi}{k}\times\\ &\times i\epsilon^i_{jk}(\delta^2(x,x_1)\widehat{J}^k_{\rho_{1}}\otimes \widehat{J}^j_{\rho_{2}}+\widehat{J}^j_{\rho_{1}}\otimes \widehat{J}^k_{\rho_{2}}\delta^2(x,x_2))+\mathcal{O}(k^{-3}).\nonumber
\end{align}
A local stationary observer who resides on the node in Fig.(\ref{punctureddisc2}) at proper distance $\ell$ to the horizon will be able to discern braided from unbraided states e.g. by measuring differences in the expectation values of the field strength operator. When considering large black holes the effect of the braiding onto the field strength would be negligible but it would become relevant for smaller (and smaller becoming) black holes.\newline
The physical picture behind the statistical phase is very similar to what happens in electromagnetism when dealing with the Aharonov-Bohm effect. To see this we use the ideas presented in \cite{Bianchi} and consider a locally flat connection on $S^2-\{p\}$
\be
A_{i a}(x)=\frac{\phi_i}{2\pi}\alpha_a(x)
\ee
with $\alpha_a(x)\in H^1(\Gamma;\mathbb{Z})$ given by
\be
\alpha_a(x)=\epsilon_{ab}\frac{x^b-x_{p}^b}{||(\vec{x}-\vec{x_p})^2||}
\ee
Ignoring the back of the sphere, its holonomy along a loop $\gamma$ is just given by
\be\label{nonabelianholonomy}
h_{\gamma}(A)=P e^{i\oint_{\gamma}\alpha_a dx^a\frac{\phi_i}{2\pi}\hat{J}^i}=e^{i n_{\gamma}\phi_i\hat{J}^i}
\ee
providing a homomorphism from $\pi_1(S^2-\{p\})$ to $SU(2)$. For the physical interpretation of the parameters $\phi^i$ one introduces the (pseudo-scalar) non-abelian magnetic field
\be B^i=\frac{1}{2}\epsilon^{ab}F_{ab}^i=\delta^2(x,x_p)\phi^i
\ee
and $\phi^i$ is just $\frac{4\pi}{k}\hat{J}^i_{\rho_{p}}$ up to a sign when using (\ref{curvaturewithpunctures}). The flux of this gravitomagnetic field through the patch $S$ of the surface of the sphere is given by
\be
\Phi^i[B,S]=\int_S B^i\epsilon_{ab}dx^a dx^b=\phi^i.
\ee
Hence, the parameters $\phi_i$ describe the flux of the gravitomagnetic field through the puncture $p$. Though it vanishes outside of $p$, one has non-local observable effects on $S^2-\{p\}$ which are captured by (\ref{nonabelianholonomy}). If we figure the boundary of the patch as $\partial S=\gamma$ then it follows by virtue of the non-abelian Stokes theorem
\be
h_{\partial S}=P e^{i\Phi_i[B,S] J^i}=P e^{i\oint_{\gamma}A}
\ee
that the exponential of the flux is just equivalent to (\ref{nonabelianholonomy}).\newline
To clarify the connection to the Aharonov-Bohm effect 
consider the superposition
\be
\tilde{\psi}=\psi_1+\psi_2=\frac{1}{\sqrt{2}}(\widehat{B}\psi+\widehat{B}^{-1}\psi)=\frac{\widehat{B}^{-1}}{\sqrt{2}}(\widehat{M}\psi+\psi)
\ee
where $\psi$ denotes the unbraided state with $||\psi||^2$=1, $\widehat{B}$ acts on the punctures $p$ and $p'$ and $\widehat{M}=\widehat{B}^2$. Then
\be\label{interference}
||\tilde{\psi}||^2=1+\Re(\langle \psi|\widehat{M}|\psi\rangle)
\ee
depends explicitly on the non-abelian phase, the interference term is proportional to $\cos(2\pi/({2+k}))$ and approaches $1$ in the large black hole limit. In contrast to the well-known situation in electromagnetism, the non-abelian phase arises here due to the Aharonov-Bohm interaction between \textit{flux-charge composites}, i.e. through the coupling of the 'non-abelian charge' $\widehat{J}^i$ of puncture/anyon $p$ and the non-abelian flux $\phi_i$ of puncture/anyon $p'$ under a complete adiabatic transport of the former around the latter. We can therefore reinterpret $\widehat{M}$ as
\be 
\widehat{M}=q^{2~\hat{J}^i_{\rho_{p}}\otimes \hat{J}^i_{\rho_{p'}}}=e^{i\hat{J}^i_{\rho_{p}}\otimes\hat{\phi}^i_{p'}}.
\ee
Hence, the phase relation between $\psi_1$ and $\psi_2$ changes under a variation of the enclosed gravitomagnetic flux $\hat{\phi}^i_{p'}$ and thus the interference pattern of (\ref{interference}) is also shifted. Since in electromagnetism the Aharonov-Bohm effect gives rise to the interpretation that the gauge potential is the true fundamental object, the measurement of the interference term by means of a local stationary observer could in principle lead to the same conclusion in the case of gravity.\newline
Before we step into the last subsection, we want to comment on the bearing of the previous discussion onto the understanding of diffeomorphism invariant states in LQG. The states of the overall quantum geometry are there given by spin networks which acquire topological degrees of freedom, i.e. knotting, through their embedding into the background manifold. Hence, the diffeomorphism invariant states fall into different knot classes whose physical relevance is usually not very well understood, despite \cite{QGSM}. The above discussion suggests, however, that - as far as the spin network close to the horizon is concerned - the physical relevance of the knotting lies in giving rise to the anyonic statistics of the horizon degrees of freedom. Furthermore, the diffeomorphism-invariant Hilbert space $\mathcal{H}_{\rm Diff}$ is not invariant under large diffeomorphisms, which form the mapping class group. We have seen in our specific case that the large diffeomorphisms do not act trivially and we can define unitary projective representations of $M_{n_{1/2},...,n_{{k/2}}}(S^2)$ on the space of diff-invariant horizon states, corresponding to the statistical phases.

\subsection{Aspects of the algebraic theory of $su(2)_k$-anyons}\label{subsectionC2}
It is known from solid state physics that quantum systems in $2d$ exhibit anyonic statistics \cite{AnyonsI, AnyonsII, MooreRead}. The question arises about how the anyonic nature of the puncture system, captured by (\ref{MonodromyMatrix},\ref{BraidingMatrix}), affects its statistics and consequently the form of its entropy. This subsection illustrates, that the Hilbert space and consequently the entropy of the IH quantum system are completely analogous to the results for a corresponding system of non-abelian anyons in condensed matter physics. We illustrate this by going through the abstract definition of a model of $su(2)_k$-anyons.\newline
The mathematical formulation of a model of non-abelian anyons in solid state physics is involved and demands more than the braid group description given above. One actually needs representations of the braid group which are compatible with the notion of fusion. The mathematical structure which consistently captures these features is a modular tensor category, specifically a unitary braided fusion category \cite{NayakTQC}. Without going into the mathematical intricacies, we will consider a particular set of classes of non-abelian anyons. These are the $su(2)_k$-anyons, which arise in non-abelian Chern-Simons theory with $G=SU(2)$ and level $k\geq 2$.\newline
A particular class of non-abelian anyons is therein defined by each value of the level $k$. For the full specification of the braiding statistics of a system of such anyons one has to give the following data:
\begin{itemize}
\item[(1.)] Anyon species/superselection sectors forming a finite set $M$: The different anyons are labeled by anyonic charges $j\in\{0,1/2,1,...,k/2\}$.
\end{itemize}
Comment: The constituents of the quantum isolated horizon form such a finite set $M$, each puncture is colored with a spin $j\in\{0,1/2,1,...,k/2\}$ and gives rise to a quantum of area $a_j=8\pi\gamma\ell_p^2\sqrt{j(j+1)}$.

\begin{itemize}
\item[(2.)] Fusion rules: Similar to ordinary spin systems, the anyon labels are combined by certain fusion rules, determining their collective behavior. For any combination of anyons $j_1, j_2, j\in M$ there is a fixed finite dimensional Hilbert space $V_j^{j_1 j_2}$ called splitting space, whereas we call $V^j_{j_1 j_2}$ the fusion space. The non-negative integers $N_{j_1 j_2}^j=\dim V_j^{j_1 j_2}=\dim V^j_{j_1 j_2}$ are called fusion multiplicities. $0\in M$ denotes the vacuum sector. In terms of the fusion matrices $N_{j}$ the composition rule reads 
\be
(j_1)\otimes (j_2) =\bigoplus_{j=|j_1-j_2|}^{\min(j_1+j_2,k-j_1-j_2)}(N_{j_{1}})_{j_{2}}^{j}~(j).
\ee
The quantum dimension $d_j$ of an anyon with charge $j$ and the fusion matrices are related by $N_j\textbf{d}=d_j\textbf{d}$. The components of the vector $\textbf{d}$ are the quantum dimensions of all anyon species occurring in the model. The total quantum dimension is defined as $\mathcal{D}\equiv\sqrt{\sum_j d_j^2}$. For $su(2)_k$-anyons the quantum dimensions are computed iteratively by $d_0=1$, $d_{1/2}=2\cos(\pi/(k+2))$ and $d_j=d_{1/2}d_{j-1/2}-d_{j-1}$ with $j\geq 1$. The Hilbert space of the $N$-punctured sphere with charges/anyons at each puncture is constructed by sewing together a chain of $(N-2)$ $3$-punctured spheres, called pants decomposition. Non-abelian anyons have $d_j>1$, which is generally not an integer. This is characteristic of the non-locality of the Hilbert space which is not simply the tensor product of $d_j$-dimensional Hilbert spaces locally associated to each anyon.
\end{itemize}
Commment:
The constituents of the quantum isolated horizon are known to obey precisely the same fusion rules. By summing over all possible puncture configurations and equipped with an appropriate combinatorial pre-factor, these rules were used in \cite{KaulMajumdar,SU2,SU2rev} to obtain for the total number of microstates the expression (\ref{AnzahlMicrozustaende}).

\begin{itemize}
\item[(3.)] The $R$-matrix: This object is used to describe an exchange of two anyons $j_1,j_2$ through braiding after the splitting of anyon $j$. The description of braiding in terms of basic data is specified by the unitary action of $R$ on splitting spaces as $R^{j_1 j_2}_j:V_j^{j_1 j_2}\to V_j^{j_2 j_1}$. Unitarity implies $N^j_{j_1 j_2}=N^j_{j_2 j_1}$ and $R$'s action is diagrammatically represented as in Fig. (\ref{fig:Rmatrix}).
\begin{figure}[htbp]
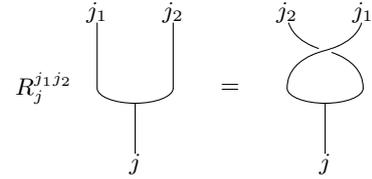

\centering
\[ \xy  
(-22,0)*{R^{j_1 j_2}_j};
(-5,0)*\ellipse(5,2)__,=:a(-180){-}; 
(7.5,0)*\ellipse(5,2)__,=:a(-180){-};
(-10,-2)*{}; (-10,-10)*{j} **\dir{-};
(-15,10)*{j_1}; (-15,0)*{} **\dir{-};
(-5,10)*{j_2}; (-5,0)*{} **\dir{-}; 
(2.5,0)*{=};
\vtwist~{(10,10)*{j_2}}{(20,10)*{j_1}}{(10,0)}{(20,0)}; 
(15,-2)*{}; (15,-10)*{j} **\dir{-};
\endxy \]
\caption{R-matrix}
\label{fig:Rmatrix}
\end{figure}
\end{itemize}
Comment:
In the context of quantum isolated horizons the $R$-matrix showed up in the discussion of the representation theory of the quantum group $U_q(su(2))$ in \cite{SU2rev}. The following point will also deal with its relation to the braiding matrix (\ref{BraidingMatrix}) and thus the statistics which we have extensively discussed in the previous subsections.

\begin{itemize}
\item[(4.)] The $F$-matrix: The fusion of three anyons is associative and therefore one has two ways to fuse three anyons to a fourth. These two ways are related by a basis change. It is specified by 
\be 
F_{j_1 j_2 j_3}^{j_4}:\bigoplus_j V^j_{j_1 j_2}\otimes V_{j j_3}^{j_4}\to \bigoplus_{j'} V^{j_4}_{j_1 j'}\otimes V_{j_2 j_3}^{j'}
\ee
and its action is diagrammatically represented as in Fig. (\ref{fig:F-matrix}).
\begin{figure}[htbp]
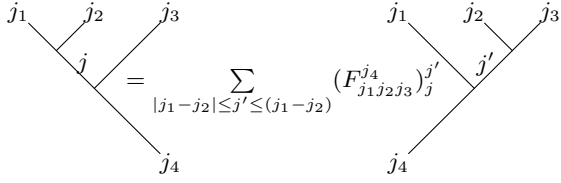

\centering
\[ \xy
(-40,10)*{j_1}; (-20,-10)*{j_4} **\dir{-};
(-30,10)*{j_2}; (-35,5)*{} **\dir{-};
(-20,10)*{j_3}; (-30,0)*{} **\dir{-}; 
(-31.5,3.25)*{j};
(-5,0)*{=\sum\limits_{|j_1-j_2|\leq j'\leq (j_1-j_2)}(F_{j_1 j_2 j_3}^{j_4})^{j'}_j};
(30,10)*{j_3}; (10,-10)*{j_4} **\dir{-};
(20,10)*{j_2}; (25,5)*{} **\dir{-};
(10,10)*{j_1}; (20,0)*{} **\dir{-};
(21.5,3.25)*{j'};
\endxy \]
\caption{F-matrix}
\label{fig:F-matrix}
\end{figure}
The $F$-matrix is unitary, obeys the orthogonality relation
\be
\sum_{l}(F_{j_1 j_2 j_3}^{j_4})^{l}_j(F_{j_4 j_1 j_2}^{j_3})^{j'}_l=\delta_{j}^{j'}
\ee
and is subject to two further consistency conditions. The first is called the pentagon relation/ Biedenharn-Elliott identity,
\be
(F^{e}_{a b h} )_{f}^{j} (F^{e}_{f c d})_{g}^{h}=\sum_{k} (F^{j}_{b c d})^{h}_{k} (F^{e}_{a k d})^{j}_{g} (F^{g}_{a b c})^{f}_{e}.
\ee
The second is termed as the hexagon relation,
\be
R^g_{ac}(F^d_{bac})^g_e R^e_{ab}=\sum_f (F^d_{bca})^g_f R^d_{af}(F^d_{abc})^f_e.
\ee
Finally, the braiding and the $R$-matrix are related by
\be\label{RBcorrespondence}
B_{j_1 j_2}=\sum_j({F_{j_1 j_3 j_2}^{j_4}}^{-1})_j^{j'} R_{j_1 j_2 }^j (F^{j_4}_{j_1 j_3 j_2})^{j'}_j,
\ee
where $B_{j_1 j_2}\in V^{j_2 j_1}_{j_1 j_2}=\bigoplus_j V_j^{j_2 j_1}\otimes V^j_{j_1 j_2}$ and its action is diagrammatically represented as in Fig. (\ref{fig:Bmatrix}).
\begin{figure}[htbp]
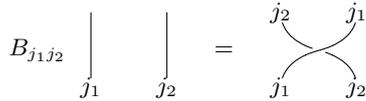

\centering
\[ \xy
(-22,5)*{B_{j_1 j_2}};
(-15,10)*{}; (-15,0)*{j_1} **\dir{-};
(-5,10)*{}; (-5,0)*{j_2} **\dir{-}; 
(2.5,5)*{=};
\vtwist~{(10,10)*{j_2}}{(20,10)*{j_1}}{(10,0)*{j_1}}{(20,0)*{j_2}}; 
\endxy \]
\caption{B-matrix}
\label{fig:Bmatrix}
\end{figure}
\end{itemize}
Comment: The $F$-matrix is the analogue of Wigner's ($q$-deformed) $\{6j\}$-symbol from recoupling theory \cite{Kauffman} which is extensively used in LQG \cite{LQG}. We identify
\be
(F_{j_1 j_2 j_3}^{j_4})_{j}^{j'}=
\begin{Bmatrix}
  j_1 & j_2 & j' \\
  j_3 & j_4 & j
\end{Bmatrix}_q.
\ee
Although the $R$-matrix has already been discussed in context of the quantum geometry of isolated horizons as in \cite{SU2rev}, its relation (\ref{RBcorrespondence}) to the braiding matrix (\ref{BraidingMatrix}) and particularly to the anyonic statistics of the model as done in the previous subsections is a novel feature. 

\begin{itemize}
\item[(5.)] The modular $S$-matrix simultaneously diagonalizes all the fusion matrices $\{N_j\}$. Through the Verlinde formula \cite{CFT} it is related to the fusion multiplicities as
\be
(N_{j_1})^j_{j_2}=\sum_d\frac{S^d_{j_2} S^d_{j_1} (S^{-1})^j_d}{S^d_0},
\ee
where $S_{j_1 j_2}=\sqrt{\frac{2}{k+2}}\sin\bigl(\frac{(2j_1+1)(2j_2+1)\pi}{k+2}\bigr)$.
\end{itemize}
Comment: In the comment to point $(2.)$ the Verlinde formula was already implicit to find the dimension of Hilbert space.  

\begin{itemize}
\item[(6.)] Topological spin $h_j$ and twist $\theta_j$:
The twist $\theta_j$ is related to the topological spin by
\be\label{topspinstatrelation}
\theta_j=e^{i2\pi h_j}=R_{j\bar{j}}^0.
\ee 
Their relation to the chiral central charge $c_{-}=c-\bar{c}$ is given by
\be\label{ccc}
\frac{1}{\mathcal{D}}\sum_j d^2_j \theta_j =e^{i \frac{2\pi}{8}c_{-}}.
\ee
For the $su(2)_k$-WZW-model used in \cite{KaulMajumdar} one has the central charge $c=3k/(k+2)$ and the conformal dimensions $h_j=j(j+1)/(k+2)$.  The topological spin feature shows up, if one considers particles in $2+1$ dimensions to be of finite extent rather than being point-like. In the context of CS-theory the thickening to a ribbon is called framing and it is needed to preserve general covariance at the quantum level \cite{Witten,Guadagnini}. Considering the possibility of a $2\pi$ rotation of a single particle relative to the rest of the system amounts to a change of the quantum wave function by a phase $e^{i2\pi\delta}$ with $\delta=h_j$. Their finite extent renders their world lines to ribbons which are twisted by such rotations. Hence, (\ref{topspinstatrelation}) expresses the \textit{(topological) spin-statistics connection} of anyons. Notice however, that $h_j$ should be differentiated from the actual spin of the object, which is related to the transformation properties with respect to the $2d$ rotation group $SO(2)$ \cite{AnyonsI}. Even if the considered system does not exhibit rotational invariance, $h_j$ is properly defined.
\end{itemize}
Comment:
When focusing on just one anyon of finite extent out of $N$ distributed on $S^2$, then it cuts out a disc $D^2$ with boundary $S^1$. In the context of anyon models one would consider $S^2$ without $D^2$ as the bulk supporting the system of the remaining $N-1$ anyons and the $1d$-circle as the edge \cite{NayakTQC}. Typically, if a $2d$-system supports anyons in the bulk, one has also chiral massless excitations propagating along the $1d$-edge described by a CFT and whose energy flux is proportional to the chiral central charge $c_{-}$. The anyons in the bulk do not determine $c_{-}$ completely, hence (\ref{ccc}) fixes $c_{-}$ only modulo $8$. Apart from this, to consider the horizon punctures as extended objects giving in turn rise to the twist $\theta_j$ is a new feature. In the large black hole limit the topological spin $h_j$ vanishes and the twist is equal to $1$, rendering the ribbon-like nature unimportant. How these qualitative considerations can have a bearing on the system of quantum gravitational anyons of the horizon and how this might be related to \cite{GhoshPranzetti} will be investigated elsewhere.\newline\
To summarize, from the points $(1.)$-$(6.)$ of this definition $(1.)$-$(3.)$ and $(5.)$ have already been known in the description of quantum isolated horizons in LQG. The latter's description for $G=SU(2)$ has been accomplished by means of a Wess-Zumino-Witten-CFT on the bounding $S^2$ \cite{KaulMajumdar}, $SU(2)_k$ CS-theory on $\mathbb{R}\times S^2$ \cite{SU2, PP} or the representation theory of the quantum deformed $SU_q(2)$, with $q$ a non-trivial root of unity \cite{SU2rev}. Not surprisingly, all these approaches led to the same expression for the dimension of the isolated horizon Hilbert space. They agree because the notions of a $2d$ modular functor, $3d$ TQFT and modular tensor category are essentially the same \cite{BakalovKirillov}. This article adds points $(4.)$ and $(6.)$ to the literature on quantum isolated horizons, especially with regard to the braiding matrix, topological spin and the twist. We want to highlight that the former are crucial in order to interpret the Hilbert space of the quantum IH as being analogous to the fusion Hilbert space of non-abelian anyons. Since the dimension of the fusion Hilbert space is computed in exactly the same manner, considering the horizon punctures as non-abelian anyons neither changes its dimension nor its entropy which both vary in $k$.\newline
Nevertheless, apart from the well known and exploited fact, that the CS-level $k$ serves as a IR cut-off by $j\leq k/2$ \cite{KaulMajumdar,SU2,SU2rev,Muxin}, from (\ref{BraidingMatrix}) we have deduced that the strength of the non-local effects due to the braiding is controlled by $k$ and that they disappear for large black holes. Hence, the non-local characteristics of the horizon Hilbert space vanish when $k\to\infty$ and we are left with the tensor product of $d_j$-dimensional Hilbert spaces that are locally associated to each horizon degree of freedom.

\subsection{$k$-dependence of the entropy and black hole radiance spectrum}\label{subsectionD2}
In the light of the previous subsections (\ref{subsectionA2},\ref{subsectionB2},\ref{subsectionC2}),  we allow ourselves to add a qualitative discussion of the relevance of the level $k$ for the entropy and the radiance spectrum of the quantum isolated horizon.\newline
Qualitatively, the braiding corresponds to non-local quantum correlations between the horizon degrees of freedom and thus adds \textit{order} to the collective. Since order reduces entropy, this suggests a reducing effect on the horizon entropy for smaller (and smaller becoming) black holes due to the correlations. If we assume without loss of generality that all punctures take $j=1/2$ in (\ref{AnzahlMicrozustaende}) then for the entropy $S\propto \log W(\{n_j\})$ one has with constant $N$ $S(k_1)<S(k_2)<S(k\to\infty)$ for levels $k_1<k_2<\infty$ and $\lim_{k\to\infty}\partial_k\bigl(S(k)\bigr)=\rm const.$. We attribute this to $k$'s double role as a cut-off and as a parameter controlling the non-local correlations. Interestingly, the analysis of the entropy $S(k)$ in \cite{Mitra} has shown that
\be\label{entropywithcorrection}
S=\lambda A +\alpha\log A,~\lambda={\rm const.}
\ee
for $k\to\infty$ (with the notorious logarithmic correction with $\alpha=-\frac{3}{2}$ as in \cite{SU2, KaulMajumdar,logcorrection}) whereas
\be\label{entropywithoutcorrection}
S=\lambda(k)A(k)
\ee
for finite $k$ as in \cite{Mitra}, i.e. small black holes. This suggests that the appearance of the logarithmic correction is related to the vanishing of the non-local effects, i.e., the collapse of the group of large diffeomorphisms to the permutation group in the large $k$-limit. We will comment on this in the discussion section.\newline
Apart from the consequences for the entropy, it could be interesting to see whether there are any traces of the non-trivial statistics of the horizon degrees of freedom in the outgoing radiation. To this aim, we invoke the following qualitative picture for the mechanism responsible for black hole radiance as given in \cite{Pranzetti,KrasnovRadiance}. Starting with the microstates given in section (\ref{subsectionB}), we assume that the black hole is initially in an eigenstate $|i\rangle$ of the horizon area operator $\widehat{A}$. Upon transition to a nearby state $|f\rangle$ with slightly smaller area, radiation of energy $\Delta E_{if}$ is emitted, which in turn leads to a reduction of the black hole energy. Let the emitted quantum be of the gravitational field with energy $\Delta E_{if}=\hbar\omega_{if}$, where $\omega_{if}$ denotes its frequency at infinity. This transition is mediated by the action of the full Hamiltonian operator on a vertex near the horizon (as on the left in Fig. (\ref{punctureddisc2})), which leads to a change in the spin associated to some of the attached edges. The evolution of the horizon from one quantum configuration to another via a dynamical phase thus corresponds to an emission (absorption) process of quanta of the gravitational field by the horizon. The analysis of the spectrum of this emission process thus yields a discrete set of lines which depend on the matrix elements of the Hamiltonian operator. For the determination of the intensities of the corresponding spectral lines and the form of the emission spectrum, one uses then the analogy to transitions in atomic physics. By virtue of Fermi's golden rule this yields for the probablity of such a transition $i\to f$
\be 
P_{if}=\frac{2\pi}{\hbar}~|\langle\widehat{H}_{if}\rangle|^2~\delta(\omega-\omega_{if})~\frac{\omega^2 d\omega d\Omega}{(2\pi\hbar)^3}.
\ee
The matrix element of the part of the Hamiltonian of the system being responsible for the transition is $\widehat{H}_{if}$ and $d\Omega$ is the differential solid angle. From this one gains the total energy $dI$ emitted by the system per unit time as
\be\label{spectrum1}
dI_{if}=2\pi\omega~p(i)~|\langle\widehat{H}_{if}\rangle|^2~\delta(\omega-\omega_{if})~\frac{\omega^2 d\omega d\Omega}{(2\pi\hbar)^3},
\ee
where $p(i)$ is the probability to find the system in the initial state $i$. Due to the fact, that the level spacing between the eigenvalues of $\widehat{A}$ decreases exponentially for large areas, the separation of the spectral lines can be rather small, thus justifying the approximation of the spectrum by a continuous profile in accordance with the calculation of the black-body spectrum derived via semi-classical arguments by Hawking.\newline
To calculate the intensity distributions, the probability distribution $p(i)$ and the matrix elements $\widehat{H}_{if}$ have to be known. These are also the relevant quantities to be inspected when checking if any (perhaps slight) alteration of the spectrum due to the braiding is expectable. Firstly, when assuming for simplicity that all accessible microstates occur with equal probability, one has $p(i)\propto e^{-S}$ since $S\propto \log W(\{n_j\})$. Knowing that $W(\{n_j\})$ is explicitly $k$-dependent and having identified that the variation of $S$ with respect to $k$ is also due to the non-local effects, i.e. the braiding, one would have $p(i)\propto e^{-S(k)}$ and the spectrum would indeed be changed by this. Secondly, the matrix elements $\widehat{H}_{if}=\langle f|\widehat{H}|i\rangle$ could very well be computed with braided states e.g. $|i'\rangle=\widehat{B}|i\rangle$ (also $\widehat{H}$ does not in general commute with the non-abelian phases) which would have a non-trivial effect on the spectrum and make it explicitly $k$-dependent. In the limit of large black holes, however, the spectrum would reduce to the one advocated in \cite{KrasnovRadiance}. We leave the issue of rigorously quantifying the spectrum in the braided case, and in the improved local setting of \cite{Pranzetti} which used the matrix elements computed in \cite{Hmatrixelements}, for future investigations.\newline
The rigorous analysis of the emission of non-gravitational quanta would require a more detailed understanding of matter couplings in LQG. Nevertheless, when invoking the semi-classical Parikh-Wilczek tunneling framework \cite{ParikhWilczek} which understands the emission of a particle from the black hole as a tunneling process, quantum gravity corrections to the emission spectrum using the entropy-area relation of (\ref{entropywithcorrection}) were given in \cite{Elias}. In the tunneling picture the emission probability is proportional to a phase space factor
\be
P_{if}\propto\frac{e^{S_f}}{e^{S_i}}=e^{\Delta S}.
\ee
Using $S$ as in (\ref{entropywithcorrection}) gives for the emission of a particle of energy $\Delta E$ from a black hole of total energy $E$
\be\label{spectrum2}
P_{if}\propto\bigl(1-\frac{\Delta E}{E}\bigr)^{2\alpha}\exp\biggl(-8\pi E \Delta E(1-\frac{\Delta E}{E}\bigr)\biggr),
\ee
which explicitly depends on the $\log$-corrections implied by LQG. For a discussion of the consequences of the first factor, see \cite{Elias}. However, when using the $k$-dependent $S$ like (\ref{entropywithoutcorrection}), the first factor drops out and $P_{if}$ becomes explicitly $k$-dependent which could provide traces of the non-trivial braiding and statistics in the outgoing radiation. A detailed analysis of these tentative arguments in the full theory would hinge much on a better understanding of the Hamiltonian operator, the matter coupling in LQG and is thus left for future investigations.

\section{Discussion and Conclusion}\label{SectionIV}
The purpose of this article was to investigate, whether and how the notion of anyonic/braiding statistics has bearing on the current LQG black hole model, based on the isolated horizon framework and the quantization as well as the symmetries of $SU(2)$-CS-theory on a punctured $S^2$. The main result is that such a model explicitly displays (non-abelian) anyonic physics (as conjectured in \cite{Sahlmann, espera, NGPF, PNG}) by direct comparison to the definition of a model of $su(2)_k$-anyons known from solid state physics in section (\ref{subsectionC2}). The non-abelian phases (\ref{KZKholonomy},\ref{MonodromyMatrix},\ref{BraidingMatrix}) responsible for the puncture exchange are in principle observable for local stationary observers. Below we discuss further implications and open questions.\newline
In quantum theories of gravity, which are based on a spacetime topology $\mathbb{R}\times M$, distinct quantum sectors labeled by the inequivalent unitary irreducible representations of the mapping class group of $M$ exist. These inequivalent quantizations, also called $\theta$-sectors, show up if the configuration space of a quantum system has a non-trivial first homotopy group \cite{CanQuantGravtheta, MathPhys}. As pointed out e.g. in \cite{Alexandrov}, the practical option in LQG is to consider the large diffeomorphisms to act trivially on the diff-invariant states \cite{LQG}. However, as we have seen above, the Hamiltonian formulation of CS-theory on the horizon gives rise to a physical Hilbert space on which unitary projective representations of the mapping class group act non-trivially. Apart from the parallel transport of the punctures, this action also leads to a braiding of corresponding incident bulk spin network edges which changes the overall knot class of the spin network. The amounting topology change of the graph, i.e. the different knottings in the vicinity of the horizon, correspond to the anyonic statistics of the punctures.\newline 
When considering large black holes ($k\to\infty$), we observed that the (non-abelian) representations of the braid/mapping class group of the punctured sphere on the Hilbert space reduce to those of the permutation group. Hence, the large $k$-limit effectively collapses much of the group of large diffeomorphisms/braids and reduces it to the respective symmetric group. The information about the knotting is apparently lost and one is solely left with the combinatorial information of the graph on which the horizon impinging spin network lives. We also lose the non-local character of the horizon Hilbert space ${\rm Inv}_k(\otimes_p j_p)$ then and are left with the 'ordinary' tensor product Hilbert space ${\rm Inv}(\otimes_p j_p)$. At this stage we mentioned at the end of section (\ref{subsectionD2}) that a numerical analysis with $j_p=1/2~\forall p$ suggests that $S(k<\infty)<S(k\to\infty)$. With this in mind let us briefly digress to the LQG treatment of the BTZ black hole in \cite{NGPF} where the question was raised, whether the difference in the entropy calculations for a BTZ black hole using the CFT and the LQG approach could be related to disregarding large diffeomorphisms in the LQG treatment. In the CFT approach the negative $\log$-correction is a result of the modular invariance (i.e. the invariance under large diffeomorphisms of the torus) of the partition function \cite{Carlip}. Back to the case of $4d$ LQG black holes, it could now be interesting to investigate what happens to the entropy if we posit the invariance of the functor $Z$ in Appendix (\ref{AppendixA}) under large diffeomorphisms/braids. Upon compactification of $\mathbb{R}$ to $S^1$, $Z$ gives the state-sum $Z_{\rm micro}=Tr(\hat{\rho}_{\rm micro})=W(\{n_j\})$. Such invariance seems to be fulfilled in the large black hole limit, since there $\widehat{M}$ is $1$, $\widehat{B}$ is $P_{pp'}\in S_{n_{1/2},...}(S^2)$ and the latter's effect is already accounted for in the combinatorial pre-factor as in (\ref{AnzahlMicrozustaende}). The entropy is then of the form (\ref{entropywithcorrection}) in comparison with (\ref{entropywithoutcorrection}) where the $\log$-correction is absent. This tentative statement seems to suggest that the $\log$-correction is indeed related to the large diffeomorphisms of the model as conjectured in \cite{NGPF}. The relation between the anyonic statistics/large diffeos and the $\log$-correction deserves a careful analysis and is left to future investigations. The same holds for a detailed analysis of the radiation spectra (\ref{spectrum1},\ref{spectrum2}) for finite $k$. In this light, it would also be important to understand better up to which (small) $k$-value the effective framework is at all useful.\newline
The scenario of anyonic statistics should also have bearing on other types of boundary surfaces (e.g with different topology than $S^2$ and/or obeying other boundary conditions) if we assume that on the kinematical level the basic data are given by a bulk spin-network graph $\Gamma$ piercing the boundary at a set of points. The phase space of the boundary theory would again give rise to (non-abelian) phases as in section (\ref{subsectionA2}). If the boundary conditions are again of IH-type but the boundary surface has non-trivial topology, the entropy can be computed as in \cite{IHBCothergen}. However, if we considered a problem obeying a different type of boundary conditions the statistical mechanics would be more complicated since a priori one is not allowed to use the powerful state-counting tools from CS-theory. See e.g. \cite{AnyonsI} for (non-analytic) thermodynamic calculations for anyon systems not exclusively described by this theory.\newline
We want to emphasize that in this article only gravitational d.o.f. were considered and these were treated as distinguishable. Recently, in \cite{PNG} it was shown that by introducing a holographic degeneracy factor accounting for matter d.o.f., only indistinguishability of the horizon states leads to a result consistent with semiclassical treatments. This is not in contradiction to this article, since we excluded matter d.o.f. from the very beginning. Upon their inclusion, one would have to use the (symmetrized) configuration space for $N$ indistinguishable objects in Appendix (\ref{AppendixC}). This would lead to a braided statistics for the horizon d.o.f. based on considering only $B_N$ and not $B_{n_{1/2},...,n_{k/2}}$. Whether a complete discussion of these points is in accordance with the suggestions of \cite{PNG} regarding the effect of anyonic statistics and how the symmetrization of the boundary condition might be related to a second quantized framework of LQG \cite{GFT}, should be clarified.\newline
The discussion laid out in section (\ref{SectionIII}) is also supported from the recent rigorous attempt at providing a full intrinsic definition of a quantum horizon from within LQG in \cite{Sahlmann}. There it was shown, that the horizon states are invariant under diffeomorphisms leaving the punctures fixed and it was noted in \cite{Sahlmann, Pranzetti}, that this symmetry gets broken, if one interchanges two differently labeled punctures. It is clear by now, that such an exchange must be mediated by a large diffeomorphism, implying the statistical symmetry of this framework. In this light it could be checked, whether the algebra of observables used in \cite{Sahlmann} also admits a non-trivial quasi-triangular Hopf algebra structure allowing for the braiding symmetry \cite{AlekseevSchomerus}.\newline
Finally, we used that topological states of matter studied in solid state physics, e.g. so-called fractional quantum Hall systems, obey non-abelian braiding statistics \cite{NayakTQC}. The analogy of the LQG black hole model to such distinct solid state systems laid out in (\ref{subsectionC2}) could in principle be used to get a better understanding of the former's nature following the spirit of "analogue gravity" \cite{AnalogueGravity}. For such topologically ordered $2d$-solid state systems a universal characterization of the many-particle quantum entanglement was found. In the entanglement entropy of such systems a universal entropy reducing constant occurs. This topological entanglement entropy accounts for the correlations related to the non-local nature of the Wilson line operators \cite{TEE} and could also be studied in the context of LQG black holes.

{\bf Acknowledgments.} A. P. thanks C. Rovelli, A. Perez (CPT) and S. Hofmann (LMU) for realizing a research internship in 11/12 supported by the ERASMUS Internship Programme. We are thankful to M. Sakellariadou, D. Oriti, M. Han, D. Pranzetti, W. Wieland and J. Th\"urigen for helpful remarks on an earlier version of this work.

\begin{appendix}\label{Appendix}
\section{Atiyah's TQFT axiomatics}\label{AppendixA}
We briefly present the axiomatization of Witten's notion of a TQFT \cite{Witten} by Atiyah \cite{Atiyah} to complement the content of sections (\ref{SectionII},\ref{SectionIII}).\newline
A $(2+1)$-dimensional TQFT $(Z,V)$ over $\mathbb{C}$ consists firstly of the association of a vector space $V(\Sigma)$ over $\mathbb{C}$ to every closed oriented smooth $2$-dimensional manifold and secondly of the association of an element $Z(M)\in V(\partial M)$ to every compact oriented smooth $3$-dimensional manifold $M$. These two associations are subject to the axioms:
\begin{itemize}
\item[1.] 
$(Z,V)$ is functorial with respect to orientation preserving diffeomorphisms of $\Sigma$ and $M$: Let $\phi:\Sigma\to\Sigma'$ be such a diffeomorphism, then one associates to it a linear isomorphism $V(\phi):V(\Sigma)\to V(\Sigma')$. For a composition of $\phi$ with $\chi:\Sigma'\to\Sigma''$ one has $V(\chi\circ \phi)=V(\chi)\circ V(\phi)$. If $\phi$ extends to an orientation preserving diffeomorphism $M\to M'$ with $\partial M = \Sigma$ and $\partial M'=\Sigma'$, one has $V(\phi)(Z(M))=Z(M')$.
\item[2.]
$(Z,V)$ is involutive, i.e. $V(-\Sigma)=V(\Sigma)^{*}$.
\item[3.]
$(Z,V)$ is multiplicative.
\item[4.]
If $\Sigma=\emptyset$ then one requires $V(\emptyset)=\mathbb{C}$ and if $M=\emptyset$ then $Z(\emptyset)=1$. For generic $\Sigma$, the identity endomorphism of $V(\Sigma)$ reads: $Z(\Sigma\times \mathbb{I}) = id_V(\Sigma)$ and crucially $\dim(V(\Sigma))={\rm Tr} V({\rm id}|_{V(\Sigma)})=Z(\Sigma\times S^1)$ gives the dimension of the respective TQFT-vector space.
\end{itemize}
Endowed with additional structure, $V(\Sigma)$ turns into a Hilbert space $\mathcal{H}_{\Sigma}$.\newline
\textbf{Example 1}: Let $M$ be a closed $3$-manifold with $\partial M=\emptyset$. Then $Z(M)\in V(\emptyset)=\mathbb{C}$ is a constant and hence the theory produces numerical invariants of $3$-manifolds. For the case of CS-theory, $Z(M)$ is just equal to (\ref{CSPI}). $Z_k(M)$ defines a topological invariant of the closed $3$-manifold $M$, which is termed as the quantum $G$-invariant of $M$ at level $k$. A natural class of gauge invariant observables of CS-theory not requiring a choice of metric are the Wilson loop operators. Let $L$ be an oriented link embedded in $M=S^3$ with $N$ components $\{C_i\}_{i=1..N}$, each of them colored with an irreducible representation $\rho_i$ of $G$. The expectation value of a product of Wilson loop operators $W(L)=\prod_{i=1}^N~{\rm Tr}_{\rho_i}[P~e^{i\oint_{C_i}A_i}]$ is 
\be
Z_k(M,L)=\langle W(L)\rangle = \frac{\int DA~e^{iS_{CS}[A]}W(L)}{\int DA~e^{iS_{CS}[A]}}.
\ee
Due to general covariance, this is invariant under smooth deformations of the (framed) link $L$. In $SU(2)_k$-CS-theory $\langle W(L)\rangle$ is equal to a corresponding evaluation of the Jones polynomial $J_L(q)$ with $q=e^{i\frac{2\pi}{k+2}}$ and it is a topological invariant of knot theory.\newline
\textbf{Example 2}: Let $\partial M=\Sigma\neq \emptyset$, then the axioms assign to the boundary the physical Hilbert space $\mathcal{H}_{\Sigma}$ and to the $3$-manifold $M$ the vector $Z_k(M)\in \mathcal{H}_{\Sigma}$, representing the time evolution of states.\newline
The axioms imply how to yield representations of mapping class groups ${\rm MCG}(\Sigma)$ of closed oriented surfaces $\Sigma$ from a $(2+1)$-dimensional TQFT. Let $\phi_t$ be the isotopy of an orientation preserving diffeomorphism $\Sigma\to\Sigma$, i.e. $\phi$ falls into one particular mapping class $[\phi]$. Then
\be
V(\phi)=\rho(\phi_t):\mathcal{H}_{\Sigma}\to \mathcal{H}_{\Sigma}
\ee
is homotopically invariant. It is implied that
\be\label{wohldefdarst}
\rho:{\rm MCG}(\Sigma)\to {\rm End}(\mathcal{H}_{\Sigma})
\ee
is a well-defined representation of ${\rm MCG}(\Sigma)={\rm Diff}^{+}(\Sigma)/{\rm Diff}_0(\Sigma)$, which acts as a symmetry on $\mathcal{H}_{\Sigma}$.\newline
The axioms also imply how to obtain the dimension of $\mathcal{H}_{\Sigma}$ for $M\cong\Sigma\times S^1$. Coupling such a TQFT to a $1$-dimensional one, corresponds to puncturing $\Sigma$ at the set of points $\{p_i\}$ by unknotted parallel circles colored with their respective representations $\{\rho_i\}$. For $\Sigma=S^2$ the application of the partition function to this configuration gives
\be\label{dimensionpuncturedsurface}
Z(S^2\times S^1;\{\rho_i\})=\dim(\mathcal{H}_{S^2;\{\rho_i\}})
\ee
and using techniques from CFT \cite{Witten,SU2Character,CFT} for a configuration of distinguishable punctures with occupation numbers $\{n_j\}$ one yields eq. (\ref{AnzahlMicrozustaende}) \cite{KaulMajumdar,SU2,SU2rev}.\newline
(\ref{wohldefdarst}) holds also for the case of punctured surfaces \cite{Witten, CFT}. The precise form of the mapping class group of the punctured sphere is recovered below.

\section{Braid group, symmetric group, pure braid group and their relations}\label{AppendixB}
\begin{figure}[htbp]
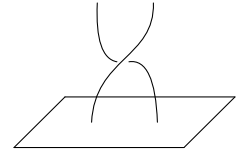

\centering
\[\xy
/r1.25pc/:{\xypolygon4"B"{~:{(3,0):(.3,.3)::}}},
"B1";"B4"**{}?<>(.5)="R1",
"B2";"B3"**{}?<>(.5)="L1",
"R1";"L1"**{}?<>(.33)="R2",
"R1";"L1"**{}?<>(.66)="L2",
"R2"+(0,3)="R3",
"L2"+(0,3)="L3",								
"L1"+(1.3,0);"R3"**@{}?!{"R1"-(1.3,0);"L3"**@{}}="M",				
"L1"+(1.3,0);"R3"**\crv{"M"-(0.75,0)&"M"+(0.75,0)},
"R1"-(1.3,0);"M"+(.1,-.1)**\crv{"M"+(0.75,0)},
"L3";"M"-(.2,.1)**\crv{"M"-(0.75,0)},
\endxy\]
\caption{Depiction of a braid.}
\label{fig:Intro Braid}
\end{figure}

Following \cite{FarbMargalit,Birman,Braiding}, facts about the braid group are gathered.
\begin{Definition}\label{DefinitionBraidGroup}
The (Artin) braid group $B_N$ on $N$ strands is an infinite group, which has $N-1$ generators $\sigma_i$, with $(1\leq i\leq N-1)$. The generators obey the following two relations
\begin{itemize}
\item[1.] $\sigma_i\sigma_j=\sigma_j\sigma_i,~~~|i-j|\geq 2$,
\item[2.] the Yang-Baxter-relation
\be 
\sigma_i\sigma_{i+1}\sigma_{i}=\sigma_{i+1}\sigma_{i}\sigma_{i+1},~~~i=1,2,...,N-2~.
\ee
\end{itemize}
\end{Definition}
$(\sigma_i)^{-1}$ denotes the inverse and $e$ the identity. The generator $\sigma_i$ corresponds to the braiding of the $i$-th strand with the $i+1$-th strand in an anti-clockwise direction, where no other strands are enclosed. The multiplication of the generators is geometrically understood as a concatenation of braids. Fig. (\ref{fig:Intro Braid}) depicts an elementary braid.
Taking the special case where $\sigma_i^2=e$ for $1\leq i\leq N-1$, the braid group reduces to the permutation group $S_N$, which is a finite subgroup of $B_N$. For braiding distinguishable strands, the pure braid group is introduced.
\begin{Definition}
The pure braid group $PB_N$ is a normal subgroup of $B_N$ and has a presentation (Burau) with the generators
\be 
\gamma_{i,j}=\sigma_{j-1}\sigma_{j-2}...\sigma_{i+1}\sigma_{i}^2\sigma_{i+1}^{-1}...\sigma_{j-2}^{-1}\sigma_{j-1}^{-1},
\ee
with $1\leq i < j \leq n$ and the following relations
\be
\gamma_{r,s}\gamma_{i,j}\gamma_{r,s}^{-1}=\notag
\ee
\be
\begin{cases}
\gamma_{i,j}\\\gamma_{i,s}^{-1}\gamma_{i,j}\gamma_{i,s}\\\gamma_{i,j}^{-1}\gamma_{i,r}^{-1}\gamma_{i,j}\gamma_{i,r}\gamma_{i,j}\\ \gamma_{i,s}^{-1}\gamma_{i,r}^{-1}\gamma_{i,s}\gamma_{i,r}\gamma_{i,j}\gamma_{i,r}^{-1}\gamma_{i,s}^{-1}\gamma_{i,r}\gamma_{i,s}
\end{cases}
\begin{matrix}
s<i~or~j<r\\i<j=r<s\\i<r<j=s\\i<r<j<s.
\end{matrix}
\ee
\end{Definition}
The action of the generator $\gamma_{i,j}$ is illustrated in Fig. (\ref{fig:Pure Braid}).
\begin{figure}[htbp]
\centering
\[ \xy
/r1.5pc/:{\xypolygon4"B"{~:{(3,0):(.3,.3)::}}},
"B1";"B4"**{}?<>(.5)="R1",
"B2";"B3"**{}?<>(.5)="L1",
"R1";"L1"**{}?<>(.1)="R2",
"R1";"L1"**{}?<>(.9)="L2",
"R2"+(0,3)="R3",
"L2"+(0,3)="L3",				
"L1"+(.3,0);"R3"**@{}?!{"R1"-(.3,0);"L3"**@{}}="M1",		
"L3";"M1"+(2.5,0)**\crv{"L3"-(0,1)&"M1"+(2.5,.5)}\POS?(.25)*{\hole}="T1"\POS?(.6)*{\hole}="T2"
?!{"R2";"R3"**@{}}="M2",
"M2"+(0,.1);"R3"**@{-}, "M2"-(0,.1);"R2"**@{-}\POS?(.5)*{\hole}="H",
"H";"M1"+(2.5,0)**\crv{"H"+(.1,0)&"M1"+(2.5,-.5)},
"H";"L2"**\crv{"H"-(.1,0)&"L2"+(0,.9)}\POS?(.27)*{\hole}="B2"\POS?(.71)*{\hole}="B1",
"T1";"T1"+(0,.6)**@{-},"T2";"T2"+(0,1)**@{-},
"T1";"B1"**@{-},"T2";"B2"**@{-},
"B1";"B1"-(0,.65)**@{-},"B2";"B2"-(0,.75)**@{-},
"L2"-(0,.3)*{i},"B1"-(-0.12,.9)*{i+1},"B2"-(0.25,.9)*{j},"R2"-(.15,.25)*{j+1},
(1,1.3)*{.},(0.7,1.3)*{.},(1.3,1.3)*{.};
\endxy\]
\caption{Depiction of a pure braid.}
\label{fig:Pure Braid}
\end{figure}
For pure braids the end points are kept fixed, whereas in $B_N$ they can be permuted. The kernel of the epimorphism $f:B_N\to S_N$ is $PB_N$, which can be compactly written as the short exact sequence 
\be
\{e\}\to PB_N\to B_N\to S_N\to \{e\}.
\ee

\section{Topology of configuration spaces for (in)distinguishable particles}\label{AppendixC}
Let the configuration space of one particle be denoted by $\mathcal{F}=X$. For $N$ indistinguishable particles one cannot make a distinction between points in $\mathcal{F}_N=X^N$ differing by the order of the particle coordinates. Let $x=(x_1,...,x_N)\in X^N$ and a different point $x'\in X^N$ with $x'=P(x)=(x_{P^{-1}(1)},...,x_{P^{-1}(N)})$, where $P\in S_N$. Physically equivalent configurations are thus orbits of points in $X^N$ with respect to $S_N$. The configuration space is $Q_N\equiv X^N/S_N$.\newline
More formally, let $M$ be a connected manifold of dimension $d=2$ or higher. Let $N$ be a positive integer, denoting the total particle number. Define Faddell's configuration space of a set of $N$ ordered points in $M$ to be
\be
\mathcal{F}_{N}(M)=\{(x_1,...,x_N)\in M\times ...\times M|x_i\neq x_j~for~i\neq j\}.
\ee
In the physical context the ordered points are distinguishable particles. In contrast,
\be
Q_N(M)\equiv \mathcal{F}_N(M)/S_N
\ee
is the configuration space of a set of $N$ unordered points in $M$, representing indistinguishable particles.\newline 
A particle exchange by means of an adiabatic transport in $d=2$ spatial dimensions is different from $d=3$. In $3d$ paths can be continuously deformed, whereas in $2d$ the topology of the configuration space allows for an oriented winding by an arbitrary number of times around other particles. Mathematically, these properties of the transport paths are captured by the first homotopy group of the configuration space. For indistinguishable particles it is given as:
\be
\pi_1\biggl(Q_N(M)\biggr)\cong S_N~(d=3);~B_N(M)~(d=2).
\ee
There are only two one-dimensional representations of $S_N$, namely the identical ($\sigma_i=1$) and the alternating one ($\sigma_i=-1$), giving in the corresponding quantum theory rise to bosonic and fermionic statistics. Quantum states for $N$ indistinguishable particles in $2d$ are elements of a Hilbert space which transforms unitarily under representations of $B_N$. If the wave functions are multiplets, one deals with higher-dimensional representations of $B_N$. These depict non-abelian anyons, giving rise to non-abelian braiding statistics, introduced in \cite{MooreRead}. The representation
\be\label{DarstellungRho}
\rho:B_N(M)\to U(\mathcal{H}_{M;N}),
\ee
maps into the unitary transformations of the Hilbert space $\mathcal{H}_{M,N}$, being in accordance with (\ref{wohldefdarst}). An element of $B_N$ acts on states as
\be
\rho(\sigma_i)~|\psi\rangle=|\psi'\rangle.
\ee
The non-abelian character is due to
\be
[\rho(\sigma_i),\rho(\sigma_j)]\neq 0.
\ee
In contrast to the above discussion, one has for distinguishable particles/punctures
\be
\pi_1\biggl(\mathcal{F}_N(M)\biggr)\cong PB_N~(d=2),
\ee
whereas for $d=3$ the fundamental group is just $e\in S_N$.\newline
If an $N$-particle system consists of a variety of distinct and thus distinguishable species, one has $n_j$ particles of species $j$ with $N=\sum_j^{j_{{\rm max}}} n_j$. The configuration space is
\be
Q_N=\mathcal{F}_N(M)/S_{n_1}\times\cdots\times S_{n_{j_{{\rm max}}}}
\ee
and its first homotopy group is
\be\label{fhdsb}
\pi_1(Q_N)=B_{n_1,\cdots, n_{j_{{\rm max}}}}(M).
\ee
It generalizes the braid group $B_N(M)$ to $j_{{\rm max}}$ distinguishable strand species. (\ref{fhdsb}) is an extension of $PB_{n_1+\cdots+n_{j_{{\rm max}}}}$ by $S_{n_1}\times\cdots\times S_{n_{j_{{\rm max}}}}$ and one has the short exact sequence
\be\label{SESExtensionMultiPaticle}
\{e\}\to PB_{n_1+\cdots}\to B_{n_1,\cdots , n_{j_{{\rm max}}}}\to S_{n_1}\times\cdots\to \{e\}.
\ee

\section{Spherical braid and pure braid group}\label{AppendixD}
A braid on $M=S^2$ has the following geometric picture. One can draw two spheres with different radii around the same center point. Moving a point on the first sphere to another position is kept track of by a strand, connecting both spheres.
The respective braid groups are $\pi_1(\mathcal{F}_N(S^2))=PB_N(S^2)$ and $\pi_1(Q_N(S^2))=B_N(S^2)$, respectively. The generators of $B_N(S^2)$ are those of $B_N$ supplemented by
\be\label{sphericalbraidgroupcondition}
\sigma_1\sigma_2...\sigma_{N-1}^2...\sigma_2\sigma_1=1.
\ee 
This constraint reflects that a closed loop can be continuously deformed and shrunk to a point on the back of the sphere due to its compactness \cite{FarbMargalit, Birman, Braiding}.\newline
The spherical pure braid group $PB_N(S^2)$ needs apart from the upper presentation for the $\gamma_{i,j}$'s the conditions 1.) $\gamma_{i,j}=\gamma_{j,i}$ for $i<j\leq N$, 2.) $\gamma_{i,i}=1$ and 3.) $\gamma_{i,i+1}\gamma_{i,i+2}...\gamma_{i,i+N-1}=1$ for $i\leq N$, where the indices in the latter are considered to run ${\rm mod}N$.\newline
Finally, for $j_{{\rm max}}$ species of punctures distributed on $S^2$ together with (\ref{SESExtensionMultiPaticle}) the braid group reads 
\be\label{SphericalBraidGroupkGruppen}
B_{n_1,...,n_{j_{{\rm max}}}}(S^2).
\ee

\section{Mapping class group and braid group on the sphere}\label{AppendixE}
Consider $S_{g,b,N}$ to be an oriented surface of genus $g$, with $b$ boundary components and a set of $N$ marked points/punctures in the surface, following \cite{FarbMargalit, Birman, Braiding}. ${\rm Homeo}^{+}(S_{g,b,N})$ is the group of orientation preserving self-homeomorphisms of $S_{g,b,N}$. These fix pointwise the boundary if $b>0$ and they map the set of $N$ marked points into itself. ${\rm Homeo}_{0}(S_{g,b,N})$ is its normal subgroup and its elements are isotopic to the identity. It is a fact, that homotopic homeomorphisms of the compact surface $S$ (even with a finite number of marked points) are isotopic, as long as $S$ is not the disc or the annulus. Additionally, one can improve homeomorphisms of this $S$ to diffeomorphisms.
Then isotopies are replaced by smooth isotopies. The mapping class group $M_{g,b,N}$, is defined as
\be
M_{g,b,N}\equiv\pi_0({\rm Homeo}^{+}(S_{g,b,N}))=\notag
\ee
\be
{\rm Homeo}^{+}(S_{g,b,N})/{\rm Homeo}_{0}(S_{g,b,N}).
\ee
With the given facts this can be restated as
\be
M_{g,b,N}\equiv\pi_0({\rm Diff}^{+}(S_{g,b,N}))=\notag
\ee
\be\label{MappingClassGroupRef}
{\rm Diff}^{+}(S_{g,b,N})/{\rm Diff}_{0}(S_{g,b,N}),
\ee
also denoted as "${\rm MCG}(S)$" or "$\Gamma_{g,N}$". ${\rm Diff}^{+}(S_{g,b,N})$ is the group of orientation preserving diffeomorphisms of $S_{g,b,N}$, that are the identity on the boundary and that act non-trivially on the punctures. They are also called "large diffeomorphisms". On the other hand, ${\rm Diff}_{0}(S_{g,b,N})$ is the group of small diffeomorphisms. Altogether, $M_{g,b,N}$ is the group of diffeomorphisms of $S$, which leave the set of punctures invariant, modulo isotopies, which leave the set of punctures invariant. It is the space of path components or isotopy classes of ${\rm Diff}^{+}(S_{g,b,N})$. However, this allows the diffeomorphisms in ${\rm Diff}^{+}(S_{g,b,N})$ to permute the $N$ punctures. In contrast, for an ordered set of $N$ punctures, indicated by $\widehat{N}$, one has ${\rm Diff}^{+}(S_{g,b,\widehat{N}})$. Due to the ordering, different orderings are discernible and the punctures are thus distinguishable. The respective pure mapping class group constitutes itself through the isotopy classes of diffeomorphisms, which preserve the punctures pointwise. It is defined as
\be\label{PureMappingClassGroupRef}
PM_{g,b,N}=
{\rm Diff}^{+}(S_{g,b,\widehat{N}})/{\rm Diff}_{0}(S_{g,b,\widehat{N}}).
\ee
There is a natural epimorphism $f:M_{g,b,N}\to S_N$, whose kernel is precisely  $PM_{g,b,N}$ and one is led to the short exact sequence
\be
\{e\}\to PM_{g,b,N}\to M_{g,b,N} \to S_N \to \{e\}.
\ee
Importantly, these groups are closely related to braid groups. In Appendix (\ref{AppendixC},\ref{AppendixD}) $\pi_1(Q_N(S^2))=B_N(S^2)$ was recovered. In \cite{Smale} it was shown that $\pi_1(SO(3))=\pi_1({\rm Diff}^{+}(S^2))=\mathbb{Z}_2$. When $N\geq 2$, this group maps non-trivially onto $\pi_1({\rm Diff}^{+}(S^2))$. The short exact sequence
\be
\{e\}\to \pi_1({\rm Diff}^{+}(S^2))\to \pi_1(Q_N(S^2))\to M_N({S^2})\to \{e\}
\ee
is equivalent to
\be
\{e\}\to \mathbb{Z}_2\to B_N(S^2)\to M_N(S^2)\to \{e\}.
\ee
From this one finds
\be
M_N(S^2)\cong B_N(S^2)/\mathbb{Z}_2.
\ee
$M_N(S^2)$ has the same generators as $B_N(S^2)$ but is supplemented by an additional condition generating the occurring $\mathbb{Z}_2$, namely
\be\label{sphericalmcgcondition}
[\sigma_1...\sigma_{N-1}]^N=1.
\ee
This is equivalent to $[\sigma_1...\sigma_{N-1}\sigma_1...\sigma_{N-2}...\sigma_1\sigma_2\sigma_1]^2=1$ when using the definition of $B_N$. Elements which obey (\ref{sphericalmcgcondition}) correspond to those of $B_N(S^2)$, where the $N$ strands are rotated by a $2\pi$ twist. This twist can be untangled when applying it twice, also known as Dirac's belt trick. In contrast to this, one has $M_{0,1,N}\cong B_N(D^2)(\cong B_N(\mathbb{R}^2))$ for the disc. For the pure case one has
\be
PM_N(S^2)\cong PB_N(S^2)/\mathbb{Z}_2.
\ee
Analogously to (\ref{SphericalBraidGroupkGruppen}), the generalization for $n_{j_{{\rm max}}}$-species leads to
\be\label{finalmappingclassgroup}
M_{n_1,...,n_{j_{{\rm max}}}}(S^2)\cong B_{n_1,...,n_{j_{{\rm max}}}}(S^2)/\mathbb{Z}_2.
\ee

\end{appendix}

\end{document}